\documentclass[%
 reprint,
superscriptaddress,
nofootinbib,
 amsmath,amssymb,
 aps,
floatfix,
]{revtex4-1}

\newcommand{\dee}[1]{\,\mathrm{d}#1\,}
\newcommand{\bstat}{$\mathcal{J}$-statistic}
\newcommand{\cstat}{$\mathcal{C}$-statistic}
\newcommand{\fstat}{$\mathcal{F}$-statistic}
\newcommand{\Tsft}{T_\mathrm{SFT}}
\newcommand{\Tobs}{T_\mathrm{obs}}
\newcommand{\Sth}{S_\mathrm{th}}
\newcommand{\zth}{z_\mathrm{th}}
\newcommand{\Tdrift}{T_\mathrm{drift}}
\newcommand{\fdrift}{f_\mathrm{drift}}

\DeclareMathOperator*{\argmax}{arg\,max}

\DeclareMathOperator*{\sinc}{sinc}

\newcommand{\mfd}{\texttt{Makefakedata\_v4}}
\newcommand{\dblbar}{\;\middle|\middle|\;}
\newcommand{\heff}{h^{\mathrm{eff}}_0}
\newcommand{\vect}[1]{\mathbf{#1}}
\newcommand{\falsealarm}{P_{\mathrm{a}}}
\newcommand{\falsedismissal}{P_{\mathrm{d}}}

\usepackage{placeins}
\usepackage{subcaption}
\usepackage{graphicx}

\usepackage{bm}

\begin{document}

\preprint{ABC/123-QED}

\title{Hidden Markov model tracking of continuous gravitational waves from a
binary neutron star with wandering spin. II. Binary orbital phase tracking}

\author{S. Suvorova}
\affiliation{\mbox{School of Electrical and Computer Engineering, RMIT University, Melbourne, Victoria 3000, Australia}}
\affiliation{\mbox{School of Physics, University of Melbourne, Parkville, Victoria 3010, Australia}}

\author{P. Clearwater}
\email{p.clearwater@student.unimelb.edu.au}
\email{patrick.clearwater@data61.csiro.au}
\affiliation{\mbox{School of Physics, University of Melbourne, Parkville, Victoria 3010, Australia}}
\affiliation{Data61, Commonwealth Scientific and Industrial Research Organisation, Corner Vimiera \& Pembroke Roads, Marsfield NSW 2122, Australia}
\author{A. Melatos}
\email{amelatos@unimelb.edu.au}
\author{L. Sun}
\email{lings2@student.unimelb.edu.au}
\affiliation{\mbox{School of Physics, University of Melbourne, Parkville, Victoria 3010, Australia}}
\author{W. Moran}
\affiliation{\mbox{School of Electrical and Computer Engineering, RMIT University, Melbourne, Victoria 3000, Australia}}
\author{R. J. Evans}
\affiliation{\mbox{Department of Electrical and Electronic Engineering, University of Melbourne, Parkville, Victoria 3010, Australia}}

\date{\today}

\begin{abstract}
	A hidden Markov model (HMM) scheme for tracking
	continuous-wave gravitational radiation from neutron stars in
	low-mass X-ray binaries (LMXBs) with wandering spin is extended by
	introducing a frequency-domain matched filter, called the \bstat{},
	which
	sums the signal power in orbital sidebands coherently.
	The \bstat{} is similar but not identical to the binary-modulated \fstat{}
	computed by demodulation or resampling.
	By injecting synthetic LMXB signals into Gaussian noise
	characteristic of the
	Advanced Laser Interferometer Gravitational-wave Observatory (Advanced
	LIGO),
	it is shown that the \bstat{} HMM tracker detects signals with
	characteristic wave strain
	$h_0 \geq 2 \times 10^{-26}$ in
	370 d of data from two interferometers, divided into 37 coherent blocks of
	equal length.
	When applied to data from Stage~I of the Scorpius~X-1 Mock Data Challenge
	organised by the LIGO Scientific Collaboration,
	the tracker detects all 50 closed injections ($h_0 \geq 6.84 \times
	10^{-26}$), recovering the frequency with a root-mean-square accuracy of
	$\leq 1.95\times10^{-5} \,\mathrm{Hz}$.
	Of the 50 injections, 43 (with $h_0 \geq 1.09 \times 10^{-25}$) are detected in a
	single, coherent 10-d block of data.
	The tracker employs an efficient, recursive HMM solver based on the Viterbi
	algorithm, which requires $\sim 10^5$ CPU-hours for a typical, broadband
	(0.5-kHz), LMXB search.
\end{abstract}

\maketitle

[Version 6.21]

\section{Introduction}

Continuous-wave gravitational radiation from accreting neutron stars in binary
systems is a key target of long-baseline interferometers like the Laser
Interferometer Gravitational Wave Observatory (LIGO) and Virgo in the
Advanced Detector Era \citep{2013riles}.
In particular, X-ray--emitting neutron stars in low-mass X-ray binaries (LMXBs)
are predicted to be relatively strong sources if they exist in a
state of torque balance \citep{1984wagoner,1998bildsten}. The
characteristic gravitational wave strain $h_0$ emitted by an LMXB in torque
balance
is proportional to the square
root of the X-ray flux independent of the distance to the source.
\citep{1998bildsten}
Scorpius X-1 (Sco X-1), the brightest LMXB in X-rays, is therefore the highest
priority target in this class.
Several plausible mechanisms exist for generating the mass or current quadrupole
moment required for torque balance,
ranging from thermocompositional and magnetic mountains
\citep{2000ushomirsky,2005melatospayne,2009vigelius,2011priymak} to $r$-modes
\citep{1998owen,2007bondarescu,2013bondarescu}. Even without torque balance,
the prospects of detecting LMXBs with persistent X-ray emission
are encouraging, depending on the detailed physics of deep crustal heating
\citep{2015haskell}.

A nonaxisymmetric rotor in a Keplerian orbit emits a frequency-modulated
gravitational wave signal. The orbital Doppler shift disperses the emitted
power into Fourier sidebands
separated in frequency by $P^{-1}$,
where
$P$ is the orbital period. Several strategies
have been deployed previously to process signals of this kind. The TwoSpect algorithm,
which operates on doubly-Fourier-transformed data, was used to conduct an
all-sky search for unknown binaries in data from LIGO Science Run 6
(S6) and Virgo Science Runs 2 and 3 (VSR2 and VSR3 respectively), returning
upper limits of $h_0 \lesssim 2\times10^{-24}$ for a whole sky search at $217\,\mathrm{Hz}$
and $h_0 \lesssim 1 \times 10^{-23}$ for a search of the frequency range
20--57~Hz for Sco~X-1
\citep{2014lsctwospect,2011goetz}. A fully templated version of TwoSpect,
tailored to handle
directed LMXB searches, offers substantial computational savings
\citep{2016meadors}. The sideband algorithm, which sums the power in the
orbital sidebands of the maximum-likelihood \fstat{}
semi-coherently, was used to conduct a directed search for Sco X-1 in
LIGO Science Run 5 (S5) data, returning an upper limit of $h_0 \leq
8\times10^{-25}$ at $150\,\mathrm{Hz}$
\citep{2015lscyoungsnr,2014sammut,2007messenger}. The radiometer algorithm applied to
LIGO S5 data returns a
model-independent upper limit of $5\times10^{-25}$ for the root-mean-square
wave strain across a
$0.25\,\mathrm{Hz}$ bin at $160\,\mathrm{Hz}$ at the sky
position of Sco X-1 \citep{2011lscradiometer,2006ballmer}. The
cross-correlation algorithm, which multiplies Fourier transforms in pairs
weighted by a phase with
an adjustable time lag \citep{2008dhurandhar,2011chung,2015whelan,2016sun}, and
the polynomial algorithm \citep{2010putten} have not yet been applied to
actual interferometer data in a Sco~X-1 search
but they competed in Stage~I of the Sco~X-1
Mock Data Challenge (MDC) \citep{2015messenger}, together with the TwoSpect,
sideband, and radiometer algorithms. Parameter-space metrics for
binary sources, a key ingredient for building semi-coherent
StackSlide-type search pipelines, have been derived recently \citep{2015leaci}
and used to estimate the optimal sensitivity of a general semi-coherent search
for Sco X-1 in the Advanced Detector Era.

A key challenge facing LMXB searches is that the spin frequency of the source,
and hence its gravitational wave frequency, wander stochastically.
Spin wandering, which is observed in X-ray pulsar timing experiments
\citep{1997bildsten}, is
driven by fluctuations in the hydromagnetic accretion torque
\citep{1993baykal,1993dekool,2004romanova} due to transient accretion disk
formation \citep{1988taam,1991baykal}
or disk-magnetosphere instabilities \citep{2004romanova}. It is auto-correlated
on
time-scales of days to weeks \citep{1997baykal}. Recent work by
\citet{2016suvorova} demonstrates that hidden Markov model (HMM) methods
offer a practical, computationally efficient strategy for tracking a
wandering frequency \citep{2001quinn}. HMM methods have been deployed with
success in many engineering applications, ranging from radar and sonar analysis
\citep{2003paris} to mobile telephony \citep{2002white}. They deliver
accurate estimation, when the signal-to-noise ratio (SNR) is low, but the sample
size is large \citep{2001quinn}, as is the case for continuous-wave searches
for gravitational radiation from neutron stars in binary systems. \citet{2016suvorova}
implemented and tested a HMM scheme based on a Bessel-weighted variant of the
maximum-likelihood \fstat{} and the
classic Viterbi HMM scheme \citep{2001quinn,1967viterbi}. The scheme
successfully detects synthetic, spin-wandering,
binary signals with $h_0 \gtrsim
8\times10^{-26}$ in Gaussian noise with
power spectral density $4\times 10^{-24}\,\mathrm{Hz}^{-1/2}$. It also detects
41
out of 50 signals without spin wandering in Stage I of the Sco X-1
MDC
with $h_0 \geq 1.1\times10^{-25}$, achieving
root-mean-square accuracy
$\leq 4\times 10^{-3}\,\mathrm{Hz}$ in frequency estimation.
A directed search of LIGO Observing Run 1 (O1) data in the range 60--650~Hz
with the HMM scheme reported an upper limit of $h_0 \lesssim 8 \times
10^{-25}$, and had a computational cost of $\sim 10^3$ CPU-hr
\citep{2017lschmmsearch}.

In this paper, we report on an improved version of the above HMM scheme, which
achieves better sensitivity while remaining competitive in terms of
computational cost. In
Ref.~\citep{2016suvorova},
the detection statistic at each HMM step is calculated by summing the \fstat{}
values at orbital sidebands weighted by positive coefficients
proportional to the squares of Bessel functions.
Physically this corresponds to summing the
sideband powers incoherently, i.e. neglecting the relative phases of the
sideband spectral components. In this paper, we replace the above detection
statistic with a variant, called the \bstat{}, that preserves the orbital
phase information. The rest of the analysis pipeline remains unchanged,
i.e. we solve the HMM recursively using the Viterbi algorithm as in previous
work. The \bstat{} takes as an input the initial orbital phase
(or equivalently the time of passage through
the orbit's ascending node or the epoch of inferior conjunction)
\citep{2014sammut}. This information is typically measured for LMXBs to an
accuracy of $\lesssim 10^{-2}\,\mathrm{rad}$
from contemporary and historical optical spectroscopic
data \citep{2016premachandra,2014galloway}. A refined measurement is
returned by the algorithm itself in the event of a detection.

The paper is structured as follows. In Section II, we review briefly the HMM
framework for frequency tracking and the Viterbi algorithm implemented to solve
the HMM. In Section III, we introduce the \bstat{} and show how it follows
naturally from the phase model of the source. The \bstat{} is constructed from
the same intermediate data products as the \fstat{}, leveraging existing and
thoroughly tested software infrastructure built by the LIGO Scientific
Collaboration.
The
improved HMM pipeline is tested against synthetic data with Gaussian noise in Section
IV and data from Stage~I of the Sco~X-1 MDC in Section V.

\section{Frequency tracking}
In this section we review briefly the HMM approach to frequency tracking, as
applied to continuous-wave searches (Section IIA), and the classic Viterbi
algorithm for solving the resulting HMM scheme (Section IIB). The reader is
referred to Ref.~\citep{2016suvorova} and references therein for a full description
of the method and its implementation. We copy the notation from
Ref.~\citep{2016suvorova} in what follows.

\subsection{HMM framework}
Let $f_\star(t)$ be the unknown, wandering spin frequency of the neutron star
as a function of time $t$. An HMM models the time series $f_\star(t)$ as a
sequence of random jumps between unobservable (`hidden') states, which
are themselves related probabilistically to some observable quantity (here, the
interferometer data) via a detection statistic. The objective of an HMM
analysis is to find the most likely sequence of jumps consistent with the
observations, once the transition probabilities are prescribed.

Continuous-wave searches are typically performed in the frequency domain on
interferometer data that have been packaged into short Fourier transforms
(SFTs) of duration $\Tsft = 30\,\mathrm{min}$, during which
$f_\star(t)$ remains confined to one frequency bin of width
$\Delta f_{\mathrm{SFT}} = (2\Tsft)^{-1}$.
Consecutive SFTs are combined to compute a frequency-domain
detection statistic $G(f)$. In between $\Tsft$ and the total observation
time $\Tobs$, for any particular astrophysical source, one can always calculate
$G(f)$ over an
intermediate `drift' time-scale $T_{\mathrm{drift}}$ ($\Tsft \leq
T_{\mathrm{drift}} \leq \Tobs$), such that $f_\star(t)$ remains confined within
one $G(f)$ frequency bin of width $\Delta f_{\mathrm{drift}} =
(2T_{\mathrm{drift}})^{-1}$, viz.
\begin{align}
	\left| \int_{t}^{t+\Tdrift} \dee{t'} \dot{f}_\star(t') \right|
	 < \Delta f_{\mathrm{drift}}
	 \label{eqn:integralcondition}
\end{align}
for all $t$. For example, in the published sideband search for Sco~X-1 in LIGO S5
data, 480 consecutive SFTs are combined to compute the sideband \cstat{}
$\mathcal{C}(f)$
for $\Tdrift = 10\,\mathrm{d}$, under the assumption that
$f_\star(t)$ wanders by less than
$\Delta f_{\mathrm{drift}} =
6\times10^{-7}\,\mathrm{Hz}$ during that time interval
\citep{2014sammut,2015lscyoungsnr}.

In an HMM search, we
compute
$G(f)$ for $N_T = \Tobs/\Tdrift$ blocks of data. In each block, the
discretised hidden variable $q(t) = f_\star(t)$ is constant and occupies one
of $N_{f_\star} = B/\Delta f_{\mathrm{drift}}$ discrete hidden states $\{q_1,
..., q_{N_{f_\star}}\}$, where $B =
f_{\star,\mathrm{max}} - f_{\star,\mathrm{min}}$ is the total search bandwidth. As
the HMM steps from one block to the next, $q(t)$ jumps from one discrete state to
another.
For a source in a binary, $G(f)$ depends not only on $f_\star$ but also
on the projected semimajor axis
of the binary orbit, $a_0 = a\sin i$,
and the orbital phase $\phi_a$ at a reference time $t_a$ (here the time of
passage through the ascending node).
Optical spectroscopy measures $a_0$ and $\phi_a$
to accuracies of $\sim25\%$ and $\sim 1\%$ respectively (see
Section~\ref{ssec:parameters} for further discussion).
\citep{2014galloway,2016premachandra}
Typically these resolutions are too coarse to produce a detectable peak in
$G(f)$ and hence the HMM output; see Figure~7 in Ref.~\citep{2016suvorova}.
Hence one must normally subdivide $a_0$ and $\phi_a$ more finely and
track a three-dimensional hidden state variable
$q(t) = [f_\star(t), a_0(t), \phi_a(t)]$, which can take on
$N_Q = N_{f_\star} N_{a_0} N_{\phi_a}$
possible values, where each $a_0$ ($\phi_a$) bin has width $\Delta a_0 =
2\sigma_{a_0}/N_{a_0}$ ($\Delta\phi_a = 2\sigma_{\phi_a}/N_{\phi_a}$),
and $\sigma_{a_0}$ ($\sigma_{\phi_a}$) is the one--standard-deviation error
bar on $a_0$ ($\phi_a$) from electromagnetic observations.
Under normal astrophysical conditions, $a_0$ and $\phi_a$ are constant during
the full search ($\Tobs \lesssim 1\,\mathrm{yr}$), and the three-dimensional HMM
reduces to its one-dimensional counterpart [with $q(t) = f_\star(t)$] computed
on a grid of $N_{a_0}N_{\phi_a}$ pairs $(a_0, \phi_a)$. We adopt the latter
approach, which is readily parallelisable, in this paper.

For a Markov process, the jump
probability for the time step $t_n$ to $t_{n+1}$ depends only on $q(t_n)$ and is
described by the transition probability matrix
\begin{align}
	A_{q_jq_i} = \Pr\left[q(t_{n+1}) = q_j \mid q(t_n) = q_i \right],
\end{align}
where $q_i$ and $q_j$ are single-index labels enumerating $N_Q$ discrete
states. As in Ref.~\citep{2016suvorova},
we approximate spin wandering as an unbiased random walk or Weiner
process: at every time step, $f_\star(t)$
jumps by $0$ or $\pm 1$
frequency bins with equal probability in the absence of discontinuous glitches
\citep{1997bildsten}.
As noted above, for observations with
$\Tobs \lesssim 1\,\mathrm{yr}$, much shorter than the mass transfer
time-scale ($T_{\mathrm{acc}} \sim 10^7\,\mathrm{yr}$), the orbital elements
are constant up to negligible corrections of order $\Tobs / T_{\mathrm{acc}}$,
and the HMM is effectively one-dimensional, with $q(t) = f_\star(t)$ and $N_Q
= N_{f_\star}$.
Hence the transition probabilities take the simple form
\begin{align}
	A_{q_jq_i} = \frac{1}{3}\left(
		\delta_{q_j,q_{i+1}} + \delta_{q_j,q_i}
		+\delta_{q_j,q_{i-1}} \right),
	\label{eqn:transmat}
\end{align}
where $\delta_{ij}$ symbolises the Kronecker delta. Other choices of the
weights, e.g $\frac{1}{4}$, $\frac{1}{2}$, $\frac{1}{4}$ are possible, but
testing shows there is little difference in performance. Machine learning
techniques for determining the weights from the data are also possible but are
beyond the scope of this paper \citep{1970baum}.

In a continuous-wave search, the observable state variable $o(t)$ corresponds
to the data collected during the interval $t \leq t' \leq t + \Tdrift$.
Formally it is a vector, whose dimension equals the interferometer sampling
frequency multiplied by $\Tdrift$. The probability that the system is observed in state
$o(t_n)$ at time $t_n$ while it occupies the hidden state $q(t_n)$ is called
the emission probability,
\begin{align}
	L_{o_jq_i} = \mathrm{Pr}\left[o(t_n) = o_j \mid q(t_n) = q_i \right]
\end{align}
In the class of continuous-wave searches considered in this paper, $L_{o_jq_i}$
can be expressed in terms of the frequency domain detection statistic
$G(f)$ as
\begin{align}
	L_{o(t_n)q_i} \propto \exp[G(f_{\star i})],
	\label{eqn:emissionprobdfn}
\end{align}
where $G(f_{\star i})$ is the log likelihood that $f_\star(t')$ lies in the
$i$-th frequency bin $\left[f_{\star i}, f_{\star i} + \Delta f_{\mathrm{drift}}
\right]$ during the interval $t_n \leq t' \leq t_n + \Tdrift$. We derive
another
version of $G(f)$, called the \bstat{}, in Section III, which generalises the
estimator in Ref.~\citep{2016suvorova} by summing the power in
orbital sidebands coherently with respect to orbital phase.

Given an observed sequence $O = \left[ o(t_0), ..., o(t_{N_T}) \right]$,
there exist $N_Q^{N_T + 1}$ hidden sequences $Q = \left[q(t_0), ...,
q(t_{N_T}) \right]$, which can give rise to $O$. Assuming the Markov property,
each hidden sequence has probability
\begin{align}
	\mathrm{P}(Q|O) &= L_{o(t_{N_T})q(t_{N_T})} A_{q(t_{N_T})q(t_{N_T - 1})} \cdots
	 L_{o(t_1)q(t_1)}
	 \notag
	 \\
	 &\phantom{=} \times A_{q(t_1)q(t_0)}\Pi_{q(t_0)},
	 \label{eqn:probsequence}
\end{align}
where
\begin{align}
	\Pi_{q_i} = \mathrm{Pr}[q(t_0) = q_i]
\end{align}
is the prior probability of each hidden state, which we take to be uniform for
simplicity, viz.
\begin{align}
	\Pi_{q_i} = N_Q^{-1}.
\end{align}
The most probable path $Q^\star(O) = \argmax \mathrm{Pr}(Q|O)$, i.e., the path
that maximises equation~(\ref{eqn:probsequence}), represents the HMM's best
estimate of the spin history $f_\star(t)$ of the source.

\subsection{Viterbi algorithm}
Many methods exist to solve efficiently for $Q^\star(O)$; see Ref.~\citep{2001quinn}
for examples. The challenge is to prune the
$N_Q^{N_T+1}$ possible hidden sequences in an efficient way. One approach,
first proposed by \citet{1967viterbi}, takes advantage of the Markov property,
and the fact that subsequences of the optimal sequence $Q^\star(O)$
are themselves optimal, to find $Q^\star(O)$
recursively
by backtracking. At every forward step in the recursion, the Viterbi algorithm
eliminates all but $N_Q$ possible state sequences; overall its computational
cost is $(N_T+1)N_Q \ln N_Q$ \citep{2001quinn}.

At forward step $k$ ($1 \leq k \leq N_T$), we calculate and store the $N_Q$ maximum probabilities
\begin{align}
	\delta_{q_i}(t_k) = L_{o(t_k)q_i} \max_{1 \leq j \leq N_Q} \left[A_{q_i
	q_j} \delta_{q_j}(t_{k-1}) \right]
	\label{eqn:vitdefdelta}
\end{align}
and the states
\begin{align}
	\Phi_{q_i}(t_k) = \argmax_{1\leq j \leq N_Q} \left[A_{q_i q_j} \delta_{q_j}
	(t_{k-1}) \right]
\end{align}
from which each maximum probability is reached, with $1 \leq i \leq N_Q$.
The optimal path is then
reconstructed by backtracking for $0 \leq k \leq N_T-1$:
\begin{align}
	q^{\star}(t_k) = \Phi_{q^{\star}(t_{k+1})}(t_{k+1})
	\label{eqn:backtrack}
\end{align}
Hence, the Viterbi algorithm computes the maximum likelihood estimator, i.e.,
$\argmax \Pr(Q|O)$.

Detailed pseudocode for the algorithm, including the
initialisation and termination
steps, is given in Ref.~\citep{2016suvorova}, following the notation and
presentation in the textbook by \citet{2001quinn}.

\section{Matched filter}
\label{sec:matchedfilter}
The emission probability $L_{o(t)q_i}$ is computed from the frequency domain
estimator $G(f)$ according to equation~(\ref{eqn:emissionprobdfn}). Many valid
choices exist for $G(f)$, depending on computational constraints, the format of
the interferometer data, and the assumed model for the phase evolution of the
source. In this paper, we leverage the existing software infrastructure for
continuous-wave searches in the LIGO Scientific Collaboration Algorithm Library
(LAL) to build $G(f)$ out of the easy-to-use and thoroughly tested
maximum-likelihood matched filter called the \fstat{} \citep{1998jks}. We
review the \fstat{} for an isolated source without any orbital
motion in Section~\ref{ssec:mffstat}. We then describe in
Section~\ref{ssec:bstat} a method to combine
\fstat{} values at orbital sidebands \emph{coherently} --- by tracking orbital
phase --- to construct a matched filter for a binary source. The
latter version of $G(f)$, termed the \bstat{}, is compared with
incoherent
algorithms for summing orbital sidebands like the \cstat{}
\citep{2007messenger,2014sammut,2015lsccrabvela} and Bessel-weighted \fstat{}
\citep{2016suvorova} in Section~\ref{ssec:relperf}.

\subsection{Isolated source: \fstat{}}
\label{ssec:mffstat}
The gravitational wave signal from a biaxial rotor without any orbital motion can
be written in the form
\begin{align}
	h(t) =
	\sum_{i=1}^4
	A_{1i}h_{1i}(t),
	+
	A_{2i} h_{2i}(t).
	\label{eqn:ht}
\end{align}
The independent components $h_{1i}(t)$ are given by
\begin{align}
	h_{11}(t) &= a(t)\cos \Phi(t), \label{eqn:h21} \\
	h_{12}(t) &= b(t)\cos \Phi(t), \\
	h_{13}(t) &= a(t)\sin \Phi(t), \\
	h_{14}(t) &= b(t)\sin \Phi(t), \label{eqn:h24}
\end{align}
where $\Phi(t)$ is the signal phase at the detector
and $h_{2i}(t)$ is obtained from $h_{1i}(t)$ by replacing 
$\Phi(t)$ with $2\Phi(t)$
in equations~(\ref{eqn:h21})--(\ref{eqn:h24}).
In~(\ref{eqn:ht})--(\ref{eqn:h24}), $A_{1i}$ and $A_{2i}$ denote arbitrary
amplitudes specific to the source, and $a(t)$ and $b(t)$ are antenna
beam-pattern
functions defined by equations~(12) and~(13) in Ref.~\citep{1998jks}, which contain
information about the source's sky position (right ascension $\alpha$,
declination $\delta$), the Earth's rotation and the detector's
orientation.
Following equations~(18) and~(96) in Ref.~\citep{1998jks}, we split the signal
phase into three terms,
\begin{align}
	\Phi(t) = 2\pi f_\star[t + \Phi_m(t;\alpha,\delta)]
	+ \Phi_s[t;f_\star^{(k)},\alpha,\delta],
	\label{eqn:phasedef}
\end{align}
where $\Phi_m$ is a time shift produced by the diurnal and annual motions of
the detector and source relative to the Solar System barycentre (SSB), and
$\Phi_s$ is a phase shift combining the latter two effects with the intrinsic
evolution of the source in its own rest frame through the intrinsic frequency
derivatives $f_\star^{(k)} = d^k\!f_\star / dt^k$ (with $k\geq 1$).

The output from a single interferometer is given by $x(t) = h(t) + n(t)$, where
$n(t)$ denotes additive noise.
Consider the special case $A_{2i} = 0$.
If the noise is Gaussian, then the normalised log
likelihood of
measuring the time series $x(t)$ over the interval $0 \leq t \leq \Tobs$
is proportional to
\begin{align}
	\ln\Lambda'_1 = (x||h) - \tfrac{1}{2}(h||h),
	\label{eqn:dfnlambda}
\end{align}
where we define the inner product
\begin{align}
	(x||y) = \frac{2}{\Tobs} \int_0^{\Tobs} \dee{t} x(t) y(t).
	\label{eqn:innerproduct}
\end{align}
Maximising $\ln \Lambda'_1$ with respect to the four amplitudes $A_{1i}$, we arrive
at the following expression for the maximum-likelihood matched filter
known as the \fstat{},
\begin{align}
	\mathcal{F} &= D^{-1}[B(x||h_{11})^2 - 2C(x||h_{11})(x||h_{12}) +
	A(x||h_{12})^2 \label{eqn:fstat1} \notag
	\\
	&\phantom{=} + B(x||h_{13})^2 - 2C(x||h_{13})(x||h_{14})
	 + A(x||h_{14})^2 ],
\end{align}
with $A = (a||a)$, $B = (b||b)$, $C = (a||b)$ and $D = AB - C^2$.
When searching the data $x(t)$ for a gravitational wave signal,
we evaluate
$\mathcal{F}$ as a function of the source parameters, e.g., $f_\star$,
$\alpha$, $\delta$, some or all of which may not be known.
A similar, independent
maximisation procedure may be performed to solve for the amplitudes $A_{2i}$.
The result is identical to~(\ref{eqn:fstat1}), except that $h_{1i}$ is replaced
by $h_{2i}$.

In practice, LIGO continuous-wave searches often take Fourier-transformed
interferometer data as inputs. It is therefore convenient to rewrite the inner
product~(\ref{eqn:innerproduct}) in terms of the Fourier transform of $x(t)$. The
calculation is presented in detail in Section~IIID of Ref.~\citep{1998jks} and also
in Ref.~\citep{prixCFStechnicalnote}. Here we quote the result.
Let $f_0$ be the search frequency, where the \fstat{} is evaluated, which may
or may not coincide with the star's spin frequency $f_\star$.
For $f_0 \neq f_\star$
we have $\left\langle \mathcal{F}(f_0) = 0\right\rangle$
and $\left\langle |\mathcal{F}(f_0)|^2 \right\rangle \approx
S_h(f_0)\Tobs$, where $\langle \cdots \rangle$ denotes
an ensemble average over many realisations of the noise,
and $S_h(f_0)$ denoted the one-sided noise power spectral density at frequency
$f_0$.
For $f_0 =
f_\star$,
we have
$\left\langle \mathcal{F}(f_0) = 0 \right\rangle$ and
$\left\langle |\mathcal{F}(f_0)|^2 \right\rangle
\gtrsim h_0^2\Tobs^2$.
Define the Fourier integral
\begin{align}
	\mathcal{F}_{1a} = \int_0^{\Tobs} \dee{t_b} x[t(t_b)] a[t(t_b)]
	e^{-i\Phi_s[t(t_b)]} e^{-2 \pi i f_0 t_b},
	\label{eqn:f2a}
\end{align}
and define $\mathcal{F}_{1b}$ in the same way but with $a[t(t_b)]$ replaced by
$b[t(t_b)]$. We
can then rewrite~(\ref{eqn:fstat1}) as
\begin{align}
	\mathcal{F} = \frac{4}{S_h(f_0)\Tobs D} \left[ B|\mathcal{F}_{1a}|^2
	- 2C \mathrm{Re}(\mathcal{F}_{1a}\mathcal{F}_{1b}^\star)
	+ A|\mathcal{F}_{1b}|^2 \right]
	\label{eqn:fstat2}
\end{align}
after rescaling by a factor $\Tobs / S_h(f_0)$ as in equation~(56) in
Ref.~\citep{1998jks}.
In~(\ref{eqn:f2a}), $t_b = t + \Phi_m(t)$ denotes a new barycentered time
coordinate related
implicitly to $t$ through the time shift introduced by the Earth's
rotation and revolution.

Formally, equation~(\ref{eqn:f2a}) integrates all the data, implying a
Fourier transform with $\sim 10^{10}$ points for $\Tobs = 1\,\mathrm{yr}$ and
kilohertz sampling.
In practice, to assist with storage,
the integral is subdivided into `atoms' \citep{prixCFStechnicalnote}.
Each atom corresponds to one SFT and is labelled by $X\alpha$, where $X$
indexes the interferometer, and $\alpha$ is the ordinal of the SFT for
that interferometer. If the SFT labelled by $X\alpha$ runs over the interval $t_{X\alpha}
\leq t \leq t_{X\alpha} + \Tsft$, equation~(\ref{eqn:f2a}) simplifies to
\begin{align}
	\mathcal{F}_{1a} = \sum_{X\alpha} \hat{a}_{X\alpha}
	\int_{t_{X\alpha}}^{t_{X\alpha}+\Tsft} \dee{t_b} x[t(t_b)]
	e^{-i \Phi_s[t(t_b)]} e^{-2 \pi i f_0 t_b}
	\label{eqn:f2aapprox}
\end{align}
with
\begin{align}
	\hat{a}_{X\alpha} = a[t(t_b = t_{X\alpha} + \Tsft / 2)].
	\label{eqn:ahat}
\end{align}
We make the approximation in equations~(\ref{eqn:f2aapprox})
and~(\ref{eqn:ahat}) that $a(t)$, which has a 24-hr period, changes slowly
during the 30-min SFT (typically without switching sign) and can be
approximated by its midpoint value.
In order to convert an SFT
(frequency bin width $\Delta f_{\mathrm{SFT}}$) into atom-based
quantities like
$\mathcal{F}_{1a}$, $\mathcal{F}_{1b}$ and $\mathcal{F}$
(frequency bin width $\Delta f_{\mathrm{drift}} =
\Tsft \Delta f_{\mathrm{SFT}} / \Tdrift \ll \Delta f_{\mathrm{SFT}}$), we `fill
in' the intermediate bins
according to the
Williams-Schutz approximation
by convolving with
the sinc function associated with the Fourier transform of
the window $t_{X\alpha} \leq t
\leq t_{X\alpha} + \Tsft$.
The reader is referred to Section 4.2 of Ref.~\citep{prixCFStechnicalnote} for full
details.

\subsection{Binary source: \bstat{}}
\label{ssec:bstat}

The gravitational wave signal from a biaxial rotor in a Keplerian orbit is
given by equations (\ref{eqn:ht})--(\ref{eqn:h24}), as for an isolated
source, except that the observed frequency is modulated by the orbital Doppler
shift, and the phase varies harmonically as
\begin{align}
	\Phi_s(t) = - 2\pi f_\star a_0 \sin \Omega(t-t_a),
	\label{eqn:phasemodel}
\end{align}
where $a_0$ is the projected semimajor axis,
$\Omega = 2\pi / P$ is the orbital angular velocity, $P$ is the orbital
period,
and $t_a = \phi_a/\Omega$ is a reference time, usually taken to be the time of passage
through the ascending node. The phase model~(\ref{eqn:phasemodel}) assumes a
circular orbit for simplicity; a nonzero orbital eccentricity is
straightforward to include in the fashion described in Section~4.5 of
Ref.~\citep{2014sammut}. Intrinsic, nonorbital frequency derivatives
$f_\star^{(k)}$ are also
omitted from~(\ref{eqn:phasemodel}) but are implemented as options in
the LAL \fstat{} code and can be activated easily via a software
switch.

Upon substituting~(\ref{eqn:phasemodel}) with $f_\star$ replaced by $f_0$
into~(\ref{eqn:f2aapprox}) and
expanding the factor $e^{-i \Phi_s[t(t_b)]}$ with the aid of the Jacobi-Anger
identity, we obtain the Fourier integral
\begin{align}
	\mathcal{J}_{1a} &= \sum_{X\alpha} \sum^\infty_{s=-\infty}
	\hat{a}_{X\alpha} J_s(2\pi f_0 a_0) e^{-is \phi_a} \notag \\
	& \phantom{=}\times 
	\int_{t_{X\alpha}}^{t_{X\alpha} + \Tsft} \dee{t_b}
	x[t(t_b)] e^{- 2\pi i (f_0 - s/P)t_b}.
	\label{eqn:f2abstat}
\end{align}
The \bstat{} is then obtained by evaluating~(\ref{eqn:fstat2})
using~(\ref{eqn:f2abstat}) and an analogous formula for $\mathcal{J}_{1b}$. The
sum over Bessel orders is truncated to $M = 2\mathrm{ceil}(2\pi f_0 a_0) + 1$
terms, because we have $|J_s(2\pi f_0 a_0)| \ll 1$ for $|s| > 2 \pi f_0 a_0 \gg
1$.

The second line in~(\ref{eqn:f2abstat})
is the same windowed Fourier transform
calculated by the \fstat{}, with $f_0^{(k)} = 0$ for all $k \geq
1$, evaluated at $f_0 - s/P$ instead of
$f_0$. Hence we can compute the \bstat{}
using existing \fstat{} infrastructure by summing the \fstat{}
output at orbital sidebands weighted by a phase factor $\propto
e^{-is\phi_a}$. Strictly speaking, according to Ref.~\citep{1998jks},
$\Phi_s$ in~(\ref{eqn:phasedef}) is allowed to depend
on $f_0^{(k)}$ for $k \geq 1$ but not on $f_0$
itself. We may therefore elect to replace $f_0$ by its average value $\bar{f_0}$
across a narrow sub-sideband (of width $1\,\mathrm{Hz}$, say)
in the argument of $J_s$, as in previous analyses using the
\cstat{} \citep{2014sammut,2015lsccrabvela}.
It is found a posteriori that the results are nearly indistinguishable.
In the previous HMM
study involving the Bessel-weighted \fstat{}, where the data are convolved with
a Bessel filter, $f_0$ is replaced by $\bar{f_0}$ in $1$-Hz sub-bands to avoid
recalculating the filter in every one of $N_{f_\star}$ frequency bins,
realising computational savings \citep{2016suvorova}.
Accordingly, a search of a large band will be implemented as a series of
searches over overlapping, $1$-Hz sub-bands.

Long-term optical spectroscopy measures
$t_a$ to an accuracy of
$|\Delta t_a| \sim 10^{-3} P$,
which translates to $\pm 1\times10^2 \,\mathrm{s}$ for Sco X-1
and $\pm 8 \times 10^{2}\,\mathrm{s}$ for Cyg X-2 for example
\citep{2014galloway,2016premachandra}.
The orbital-phase--coherent \bstat{} is
sensitive to $t_a$ through (\ref{eqn:f2abstat}).
To preserve orbital phase coherence the condition
$2\pi\bar{f_0}a_0 \Omega |\Delta t_a| \ll
1$ must be satisfied; the absolute error $|\Delta t_a|$ contributes cumulatively
to every sideband, and there are $\approx 4\pi f_0 a_0$ significant
sidebands.
In terms of fiducial Sco~X-1 parameters, one requires
\begin{align}
	|\Delta t_a| \lesssim 4.0 (f_0 / 300\,\mathrm{Hz})^{-1} (a_0 /
	1.44\,\mathrm{s})^{-1}
	(P/68023\,\mathrm{s})\,\mathrm{s}.
	\label{eqn:scox1target}
\end{align}
The accuracy targeted in~(\ref{eqn:scox1target}) is unachievable at the
time of writing, so we are obliged to either estimate $t_a$ or search over it.
Constraints from
electromagnetic data nevertheless reduce the search domain significantly.
In this paper, we elect to search over $t_a$ (or equivalently $\phi_a$).
The results are presented in
Sections~\ref{sec:synthdata} and~\ref{sec:mdc}.

We note in passing that
any algorithm that
sums \fstat{} values at orbital sidebands, like the \bstat{}, \cstat{}
\citep{2007messenger,2014sammut}
and Bessel-weighted \fstat{} \citep{2016suvorova}, is not truly a
maximum-likelihood estimator.
The \fstat{}
at the $s$-th sideband (frequency $f_0 - s/P$) maximises the partial likelihood
of detecting a signal at $f_0 - s/P$
with respect to amplitudes $A^{(s)}_{1i}$ and $A^{(s)}_{2i}$ specific to that
sideband,
\emph{not} the total log likelihood $\ln \Lambda'_1$ for all the
sidebands added together.
A true maximum-likelihood estimator would maximise $\ln \Lambda_1'$
for a single, optimal choice of the eight amplitudes $A_{1i}$
and $A_{2i}$.
We discuss
quasi-maximum-likelihood
estimators further in
Appendix~\ref{sec:app:qmle}.

Before presenting results based on searching over $\phi_a$, we comment briefly
on an alternative approach: estimating $\phi_a$.
Formally, $\mathcal{J}_{1a}$ in~(\ref{eqn:f2abstat})
is a Fourier series in the
variable $\phi_a$ with period $2\pi$. Therefore, upon computing the discrete
Fourier transform of $\mathcal{J}_{1\alpha}$ in the variable $s$,
we expect to observe a peak at
the true value of $\phi_a$.
The calculation is fast, but it yields multiple
spurious peaks, when the signal
approaches the detection limit.
In principle, the
peaks can be vetoed by the recursive
logic of the HMM, but $\phi_a$ is known to be constant
astrophysically on the time-scale $\Tobs$,
so the HMM
reduces
equivalently to searching over $\phi_a$ without Fourier maximisation.

\subsection{Relative performance}
\label{ssec:relperf}

Before combining the \bstat{} with the HMM in Sections~\ref{sec:synthdata}
and~\ref{sec:mdc}, we compare its sensitivity with matched filters used
in previous work.
The comparison is based on injections into a single, 10-d block of data,
typical of a single HMM detection step in a Sco X-1 search ($\Tdrift =
10\,\mathrm{d}$), with noise level $S_h(f_\star)^{1/2} = 4 \times
10^{-24}\,\mathrm{Hz}^{1/2}$ and other injection parameters as in
Table~\ref{tbl:synthparams}.

\begin{figure*}
	\begin{subfigure}[b]{0.45\textwidth}
		\includegraphics[width=\textwidth]{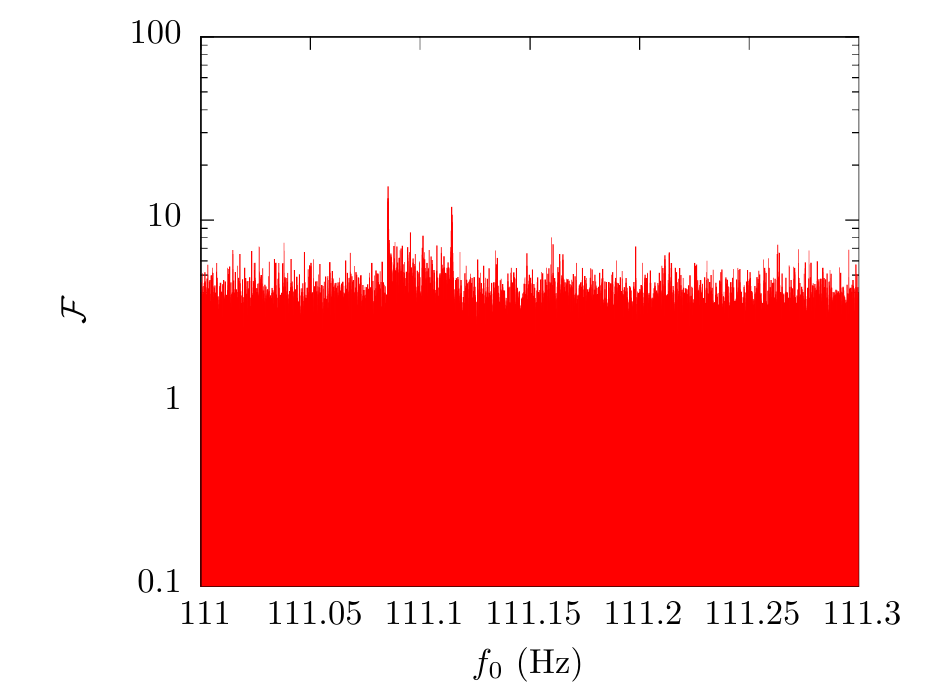}
		\caption{}
		\label{fig:comp:fstat}
	\end{subfigure}
	\hfill
	\begin{subfigure}[b]{0.45\textwidth}
		\includegraphics[width=\textwidth]{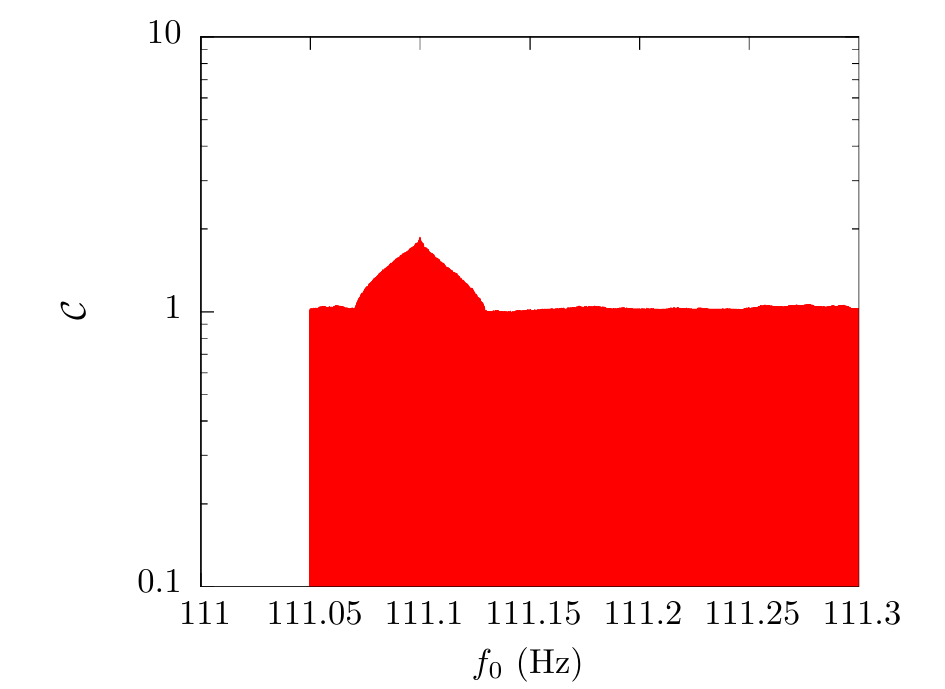}
		\caption{}
		\label{fig:comp:cstat}
	\end{subfigure}
	\begin{subfigure}[b]{0.45\textwidth}
		\includegraphics[width=\textwidth]{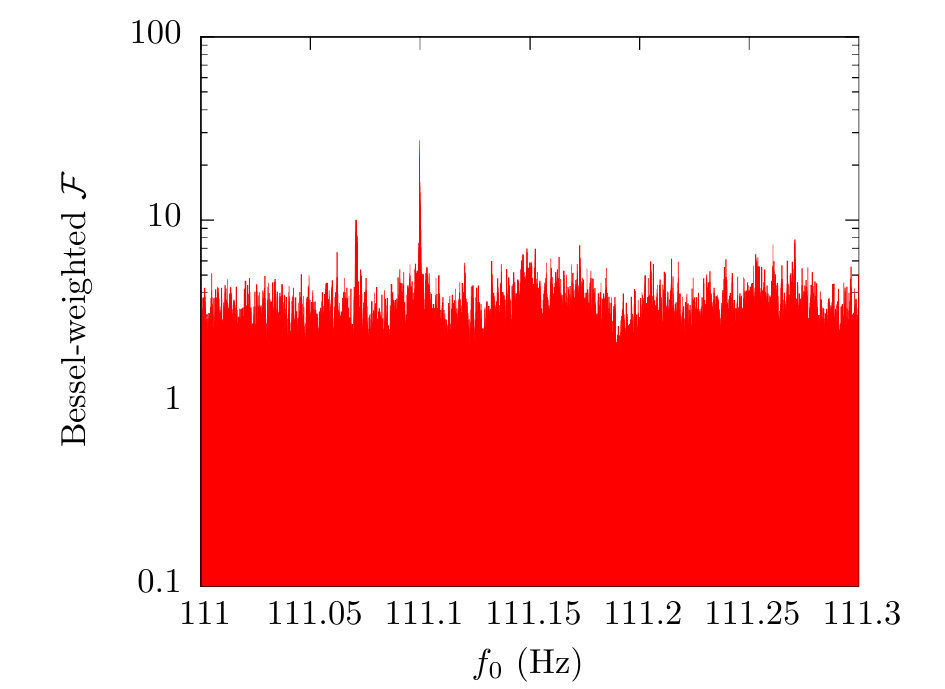}
		\caption{}
		\label{fig:comp:besselwt}
	\end{subfigure}
	\hfill
	\begin{subfigure}[b]{0.45\textwidth}
		\includegraphics[width=\textwidth]{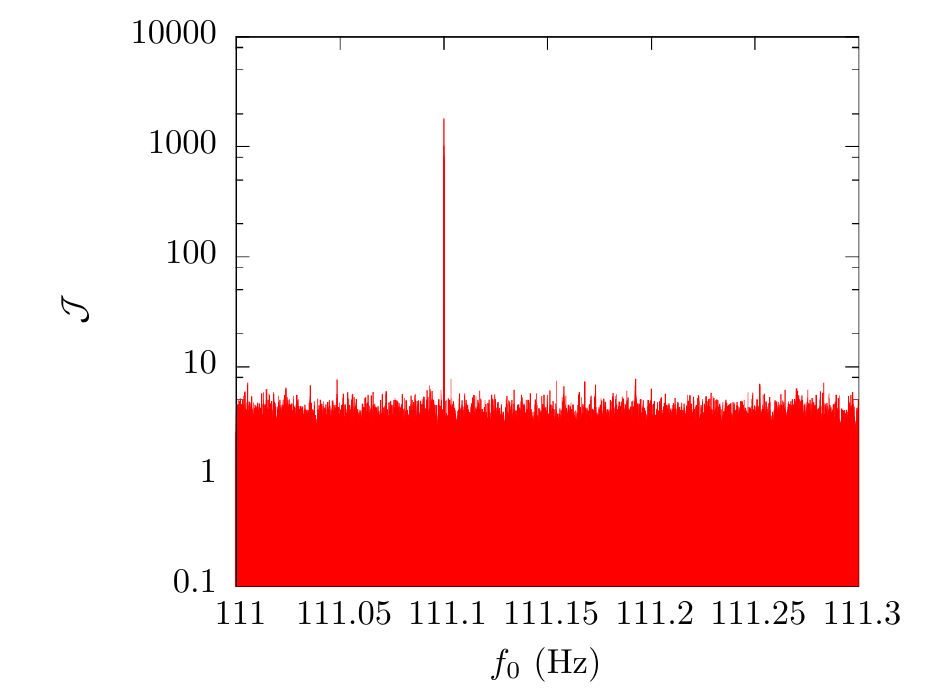}
		\caption{}
		\label{fig:comp:bstat}
	\end{subfigure}
	\caption{Signature of an injected binary signal into the
	(a)~\fstat{} \citep{1998jks},
	(b)~\cstat{} \citep{2014sammut},
	(c)~Bessel-weighted \fstat{} \citep{2016suvorova} and
	(d)~\bstat{}. The search frequency $f_0$ is plotted on
	the horizontal axis (units: Hz).
	All plots are generated from the same synthetic data.
	The plots are normalised so that the mean of the noise is unity; the mean
	of the noise appears to be above unity in panels
	(a) and (d) because each pixel
	represents many bins.
	Signal power leaks into orbital sidebands in
	panels~(a)--(c) but is
	concentrated in a single frequency bin in panel~(d).
	Injection parameters are the same
	as in Table~\ref{tbl:synthparams} with $S_h(f_\star)^{1/2} =
	4\times10^{-24}\,\mathrm{Hz}^{1/2}$ and $h_0 = 8 \times
	10^{-25}$.
	}
	\label{fig:comparison}
\end{figure*}

Figure~\ref{fig:comparison} displays the output of four matched filters as a function of $f_0$:
the \fstat{} \citep{1998jks}, \cstat{} \cite{2007messenger,2014sammut},
Bessel-weighted \fstat{} \citep{2016suvorova}, and \bstat{}
(Section~\ref{ssec:bstat} of this paper). The injected signal is
strong, with $h_0 = 8\times10^{-25}$, making it visible to the eye in all four
panels.
The matched filters are evaluated for the exact,
injected values of $a_0$ and $\phi_a$, i.e., $f_0$ is the only search
parameter. In Figure~\ref{fig:comp:fstat} we see the distinctive double-horn profile of a
binary source in the \fstat{}, which arises because a source with
inclination angle $\iota \neq 0$
spends more time moving parallel to the line of sight (maximum Doppler
shift) than perpendicular to it. (The Fourier transform is the
time-weighted frequency histogram.) In Figure~\ref{fig:comp:cstat} we see the distinctive
onion-dome profile of the \cstat{} output. The peak is centred
on the
value of $f_0$
bracketed by the maximum number of
significant sidebands. It is broad, because sidebands
are summed with equal weights; for weaker but still detectable
signals, the peak shrinks and
merges into a flat plateau raised above the noise.
In Figure~\ref{fig:comp:besselwt}
the onion done transforms into a concave cusp. The peak is sharper
and taller than in Figure~\ref{fig:comp:cstat}, because the Bessel weighting
favours the central sidebands, which are intrinsically stronger.
In Figure~\ref{fig:comp:bstat}, corresponding to the \bstat{}, the peak is
even taller. Essentially zero power falls outside the central bin in the
\bstat{};
by
accounting for the phases $\propto e^{-i s \phi_a}$ in~(\ref{eqn:f2abstat}),
we avoid power leaking into the shoulders of the peak
(cf. Figure~\ref{fig:comp:besselwt}).
The peaks are 11.8, 2.7, 14.6, and 32.6 dB above the
noise (that is, the mean for bins containing only noise) in Figures \ref{fig:comp:fstat},
\ref{fig:comp:cstat}, \ref{fig:comp:besselwt}, and \ref{fig:comp:bstat}
respectively (note logarithmic vertical scale).

\begin{figure*}
	\begin{subfigure}[b]{0.49\textwidth}
		\includegraphics[width=\textwidth]{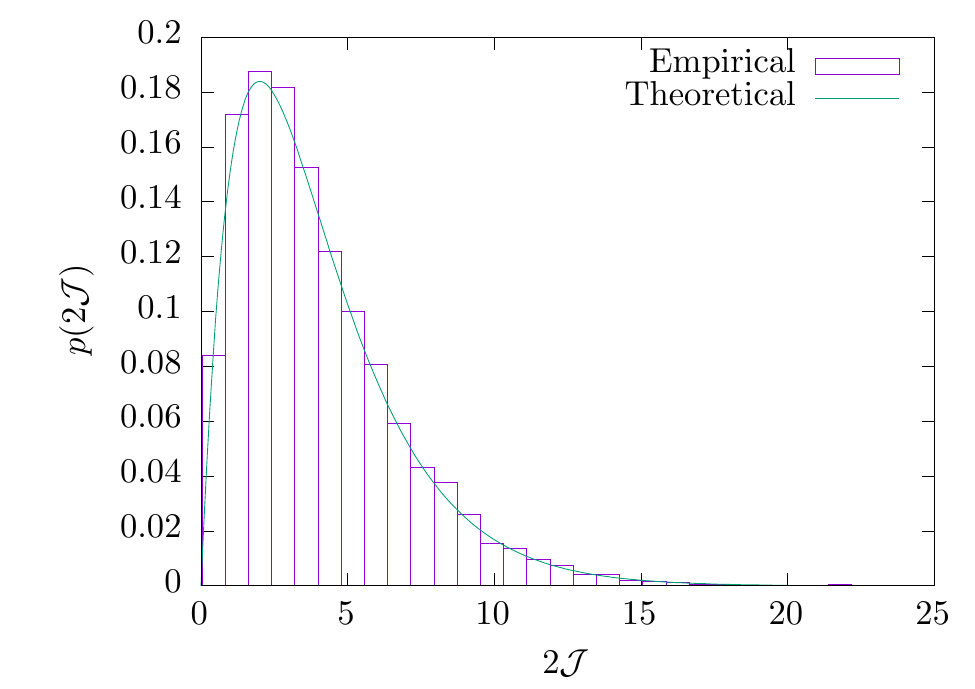}
		\caption{}
		\label{sfig:2:noisepdf}
	\end{subfigure}
	\hfill
	\begin{subfigure}[b]{0.49\textwidth}
		\includegraphics[width=\textwidth]{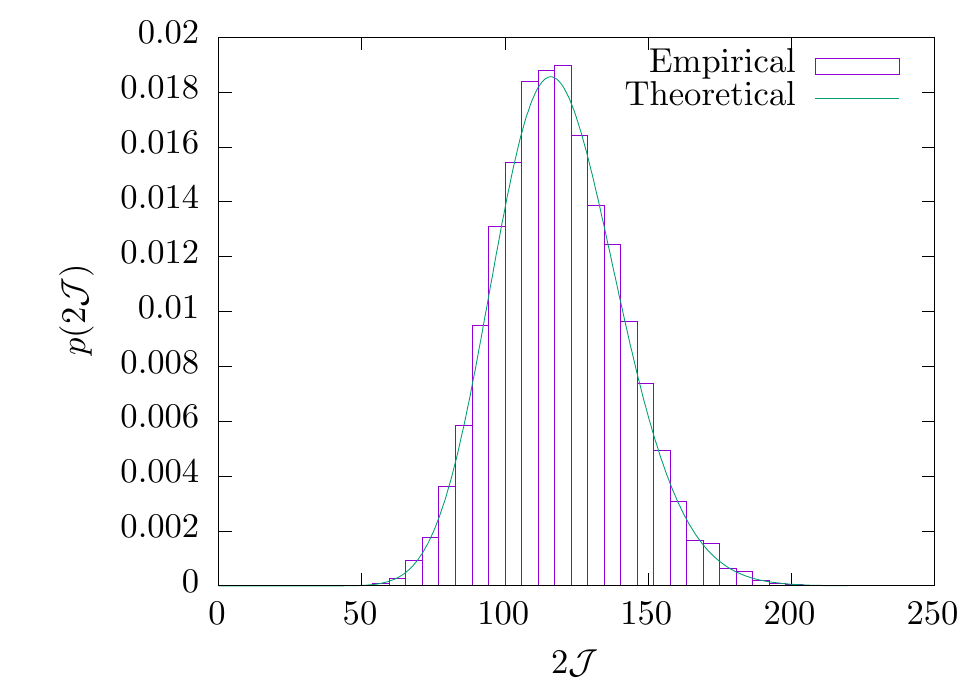}
		\caption{}
		\label{sfig:2:signalpdf}
	\end{subfigure}
	\caption{Probability density function of the \bstat{}.
	(a) Noise only. (b) Noise plus
	injected binary signal with $h_0 = 8\times10^{-26}$ and other parameters as in
	Table~\ref{tbl:synthparams} including $S_h(f_\star)^{1/2} =
	4\times10^{-24}\,\mathrm{Hz}^{1/2}$, as in Figure~\ref{fig:comparison}.
	The empirical histograms (purple columns) are
	generated from $10^4$ realisations.
	The theoretical (green) curves correspond to~(\ref{sfig:2:noisepdf})
	$\chi^2(2\mathcal{J}; 4, 0)$ and~(\ref{sfig:2:signalpdf})
	$\chi^2(2\mathcal{J}; 4, \lambda)$, with $\lambda = 111$ (empirical fit).
	}
	\label{fig:bstatpdf}
	%
\end{figure*}

Figure~\ref{fig:bstatpdf} shows the probability density function (PDF) for the
\bstat{} for pure noise (Figure~\ref{sfig:2:noisepdf})
and noise plus signal (Figure~\ref{sfig:2:signalpdf}).
Like the \fstat{} \citep{1998jks},
the \bstat{} is distributed as a central
chi-squared distribution with four degrees of freedom, $\chi^2(4, 0)$ for
white, Gaussian noise (regardless of the
noise amplitude). When a signal is added to
the noise, the \bstat{} is distributed as a non-central chi-squared
distribution with four degrees of freedom, $\chi^2(4, \lambda)$.
The non-centrality parameter $\lambda$ is related to the
amplitude of the signal, the amplitude of the noise and the observation time
according to
\begin{align}
	\lambda \propto \frac{h_0^2 \Tobs}{S_h(f_0)}.
	\label{eqn:lambdadef}
\end{align}

\section{Synthetic data}
\label{sec:synthdata}
\subsection{Injection and search procedure}

To assess the effectiveness of the \bstat{},
we begin by seeking
to detect synthetic signals injected into white, Gaussian noise.
To facilitate comparison with previous work,
we re-use the trial parameters used by \citet{2016suvorova} in an identical
test of the Bessel-weighted \fstat{}.
The
parameters are quoted in
Table~\ref{tbl:synthparams}. Data are generated for
$\Tobs = 370\,\mathrm{d}$, divided into 37 blocks of length
$\Tdrift = 10\,\mathrm{d}$.
The source frequency $f_\star(t)$ is constant within each block and jumps
discontinuously by at most one frequency bin ($\Delta f_\mathrm{drift} =
5.787037 \times 10^{-7}\,\mathrm{Hz}$) up or down when passing from one block
to the next.
We generate data
for two interferometers (to facilitate comparison with Section~\ref{sec:mdc})
using the \mfd{} tool
from the
LAL software suite.

\begin{table}
	\caption{Injection parameters for the trials on synthetic data in
	Section~\ref{sec:synthdata}.}
	\label{tbl:synthparams}
	\begin{tabular}{lll}
		\hline
		Parameter & Value & Units \\
		\hline
		$f_\star$ & $111.1$ & Hz \\
		$\dot{f}_\star$ & $0$ & $\mathrm{Hz}\,\mathrm{s}^{-1}$ \\
		$\alpha$ & $4.2757$ & rad \\
		$\delta$ & $-0.27297$ & rad \\
		$\cos\iota$ & $0.71934$ & -- \\
		$\psi$ & $4.08407$ & rad \\
		$S_h(f_\star)^{1/2}$ & $4\times10^{-24}$ & $\mathrm{Hz}^{-1/2}$ \\
		\hline
		$P$ & $68023.7$ & s \\
		$a_0$ & $1.44$ & s \\
		$T_p$ & $1245984672$ & s \\
		\hline
	\end{tabular}
\end{table}

\subsection{Optimal path}
Table~\ref{tbl:synthresults} lists the outcomes of five trials
with $1.5 \leq h_0/10^{-26} \leq 8$.
It shows whether each signal is detected as the optimal Viterbi path
and quotes the
root-mean-square error $\epsilon_{f_\star}$
between the optimal path and $f_\star(t)$.
We see that the \bstat{} is
able to recover signals with
$h_0 \geq 2\times10^{-26}$,
consistent with the result in
Ref.~\citep{2016suvorova} for isolated pulsars. The
error amounts to $\epsilon_{f_\star} \sim 10^{-7}\,\mathrm{Hz}$
for all cases where there is a
detection, i.e.
as long as
the signal can be detected,
$\epsilon_{f_\star}$ does not worsen, as
$h_0$ decreases.
The error also satisfies $\epsilon_{f_\star} \lesssim \Delta\fdrift$, i.e., the error
is comparable to the frequency resolution of the \bstat{}.

\begin{figure*}
	\begin{subfigure}[b]{0.49\textwidth}
		\includegraphics[width=\textwidth]{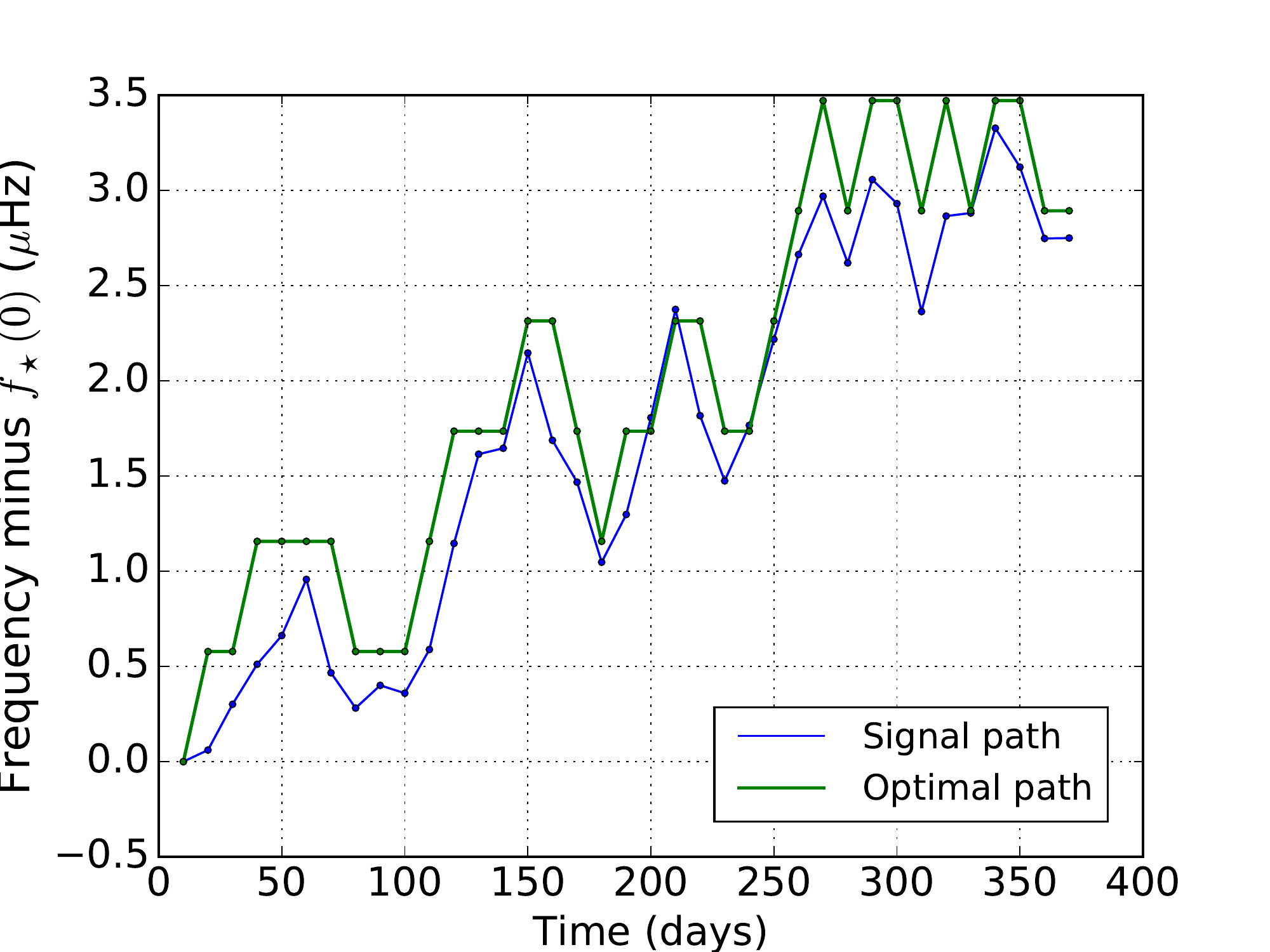}
		\caption{}
		\label{sfig:3:1}
	\end{subfigure}
	\hfill
	\begin{subfigure}[b]{0.49\textwidth}
		\includegraphics[width=\textwidth]{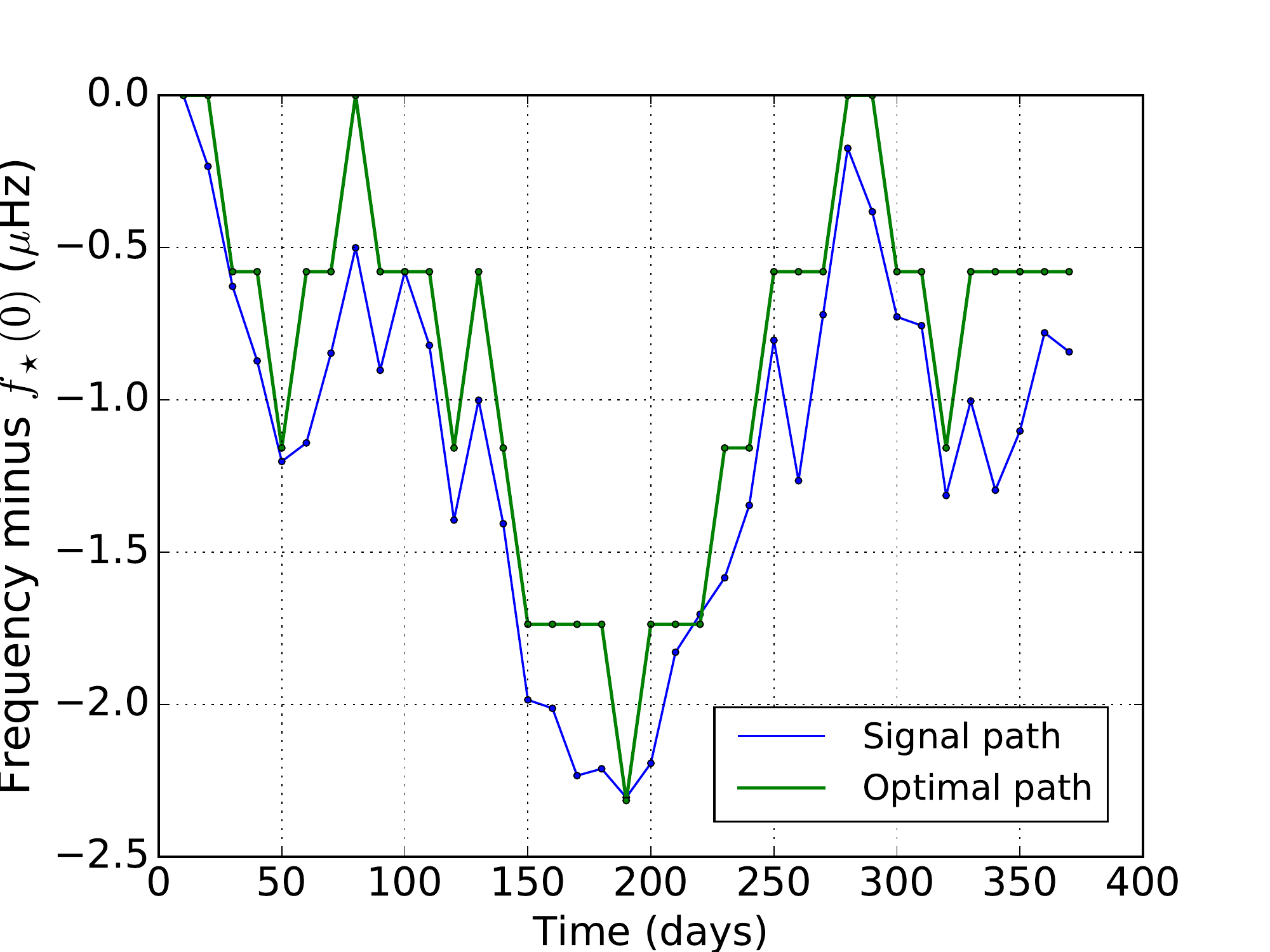}
		\caption{}
	\end{subfigure}
	\begin{subfigure}[b]{0.49\textwidth}
		\includegraphics[width=\textwidth]{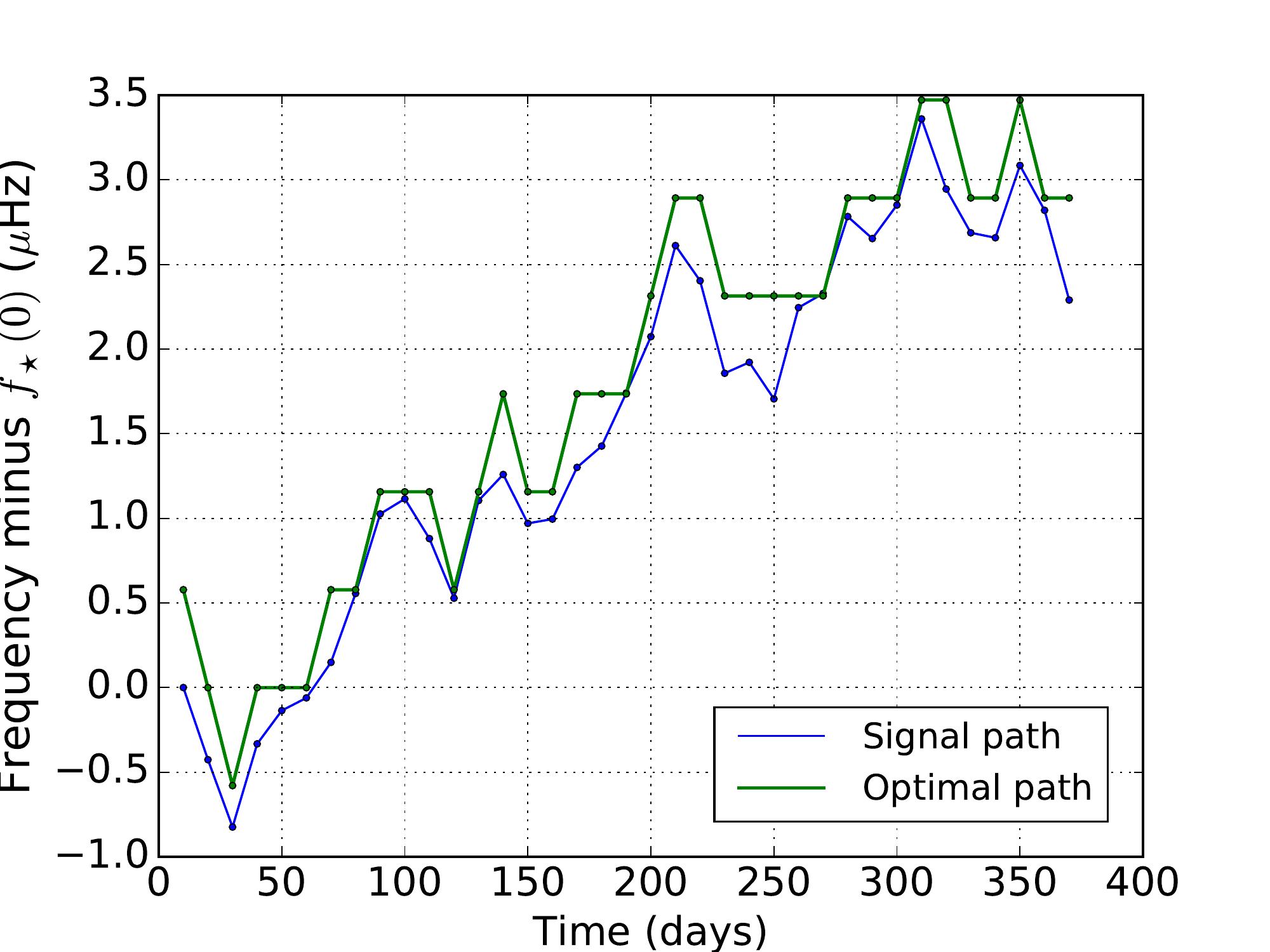}
		\caption{}
	\end{subfigure}
	\hfill
	\begin{subfigure}[b]{0.49\textwidth}
		\includegraphics[width=\textwidth]{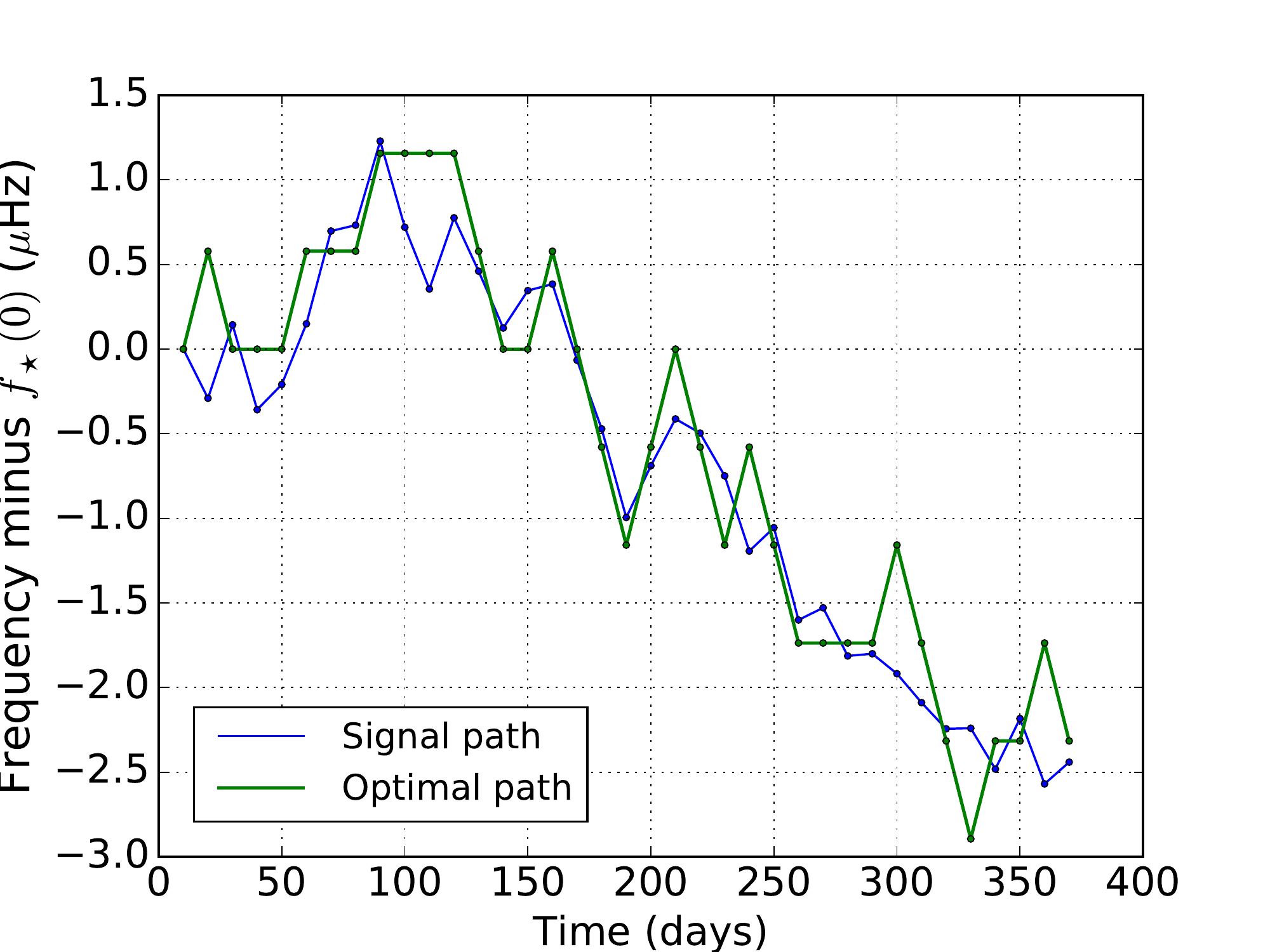}
		\caption{}
		\label{sfig:3:4}
	\end{subfigure}
	\begin{subfigure}[b]{0.49\textwidth}
		\includegraphics[width=\textwidth]{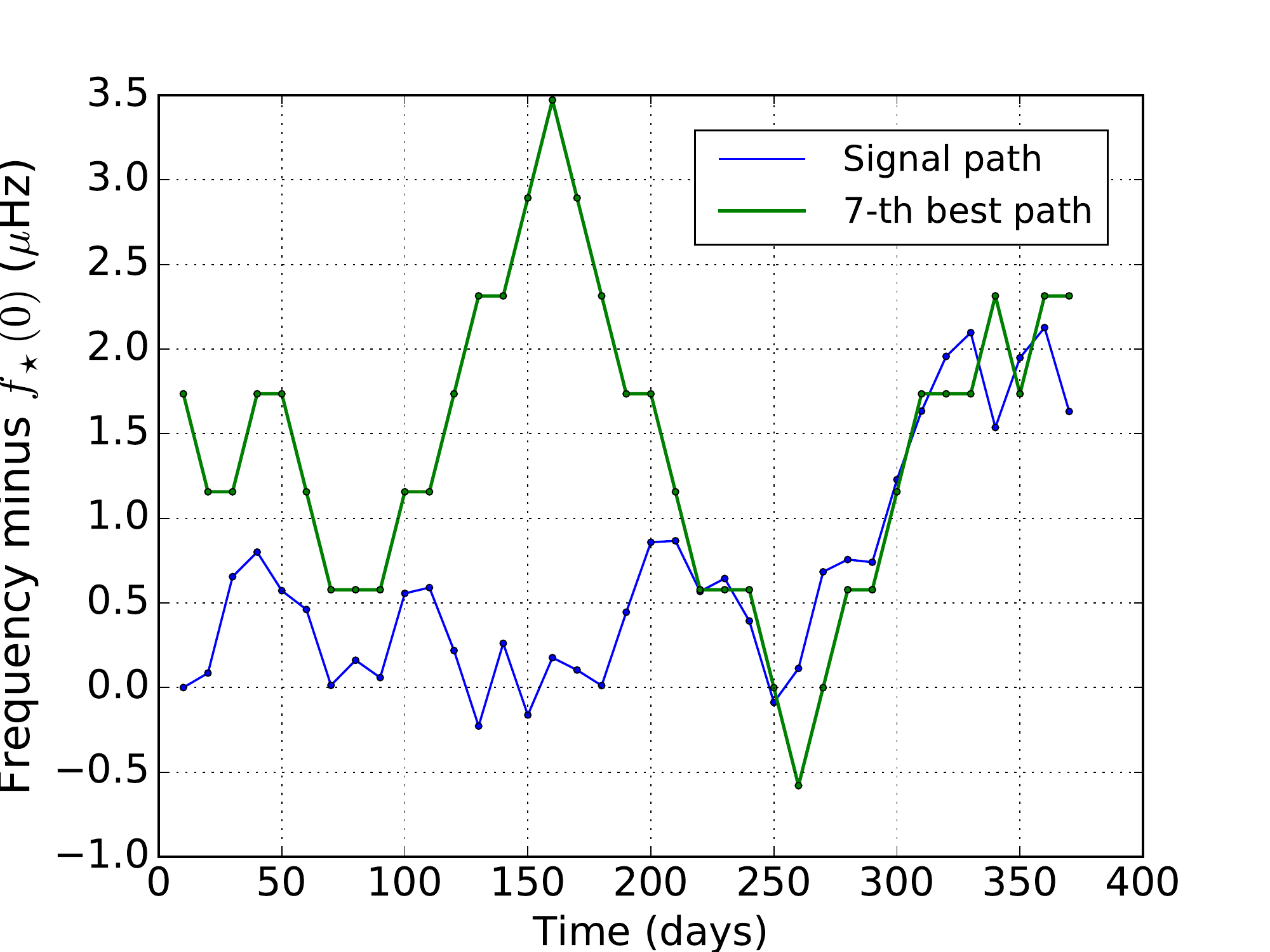}
		\caption{}
		\label{sfig:3:5}
	\end{subfigure}
	\caption{True $f_\star(t)$
	(blue curve) and
	Viterbi path (green curve) for the
	five injected signals in
	Table~\ref{tbl:synthresults} with binary parameters quoted in
	Table~\ref{tbl:synthparams}.
	Panels (a)--(d) correspond to
	$h_0/10^{-26} = 8, 5, 4, 2$ respectively; a good match between the true and
	optimal Viterbi paths is obtained in
	all cases.
	In panel~(e), with $h_0/10^{26} = 1.5$,
	the best Viterbi path
	lies outside the graph's bounding box; the seventh-best path is plotted
	instead, which matches $f_\star(t)$ best in a root-mean-square sense.
	The units on the horizontal and
	vertical axes are days and $\mu\mathrm{Hz}$ respectively.
	}
	\label{fig:paths}
\end{figure*}

Figure~\ref{fig:paths} overplots $f_\star(t)$
against the paths recovered by the Viterbi algorithm. For signals
with $h_0 \geq 2.0\times 10^{-26}$, the optimal path returned by Viterbi
closely matches
$f_\star(t)$ as noted above.
There is a slight mismatch of order one \bstat{} frequency bin,
because $f_\star(t)$ wanders continuously,
whereas the
HMM transitions between discrete
bins. For $h_0 = 1.5\times 10^{-26}$, just below the
detection threshold,
the optimal path is $0.4\,\mathrm{Hz}$ from
the injected path, outside the range plotted in Figure~\ref{sfig:3:5}. Instead
Figure~\ref{sfig:3:5} shows the seventh-ranked Viterbi path,
which minimises $\epsilon_{f_\star}$.
The latter path deviates from $f_\star(t)$ in the first half of the data
but recovers to converge on
$f_\star(t)$ towards the end.

The strongest injection in Figure~\ref{sfig:3:1}, with $h_0 = 8.0
\times 10^{-26}$, matches the weakest signal
detected by the Bessel-weighted \fstat{}
\citep{2016suvorova}. With the orbital phase now taken into account,
the \bstat{}
detects the signal without difficulty.
Going further, the \bstat{} detects injections down to
$h_0 = 2.0 \times 10^{-26}$.
A signal with $h_0 = 2.0 \times 10^{-26}$
corresponds to the weakest
isolated source
(zero orbital motion) detected
in Ref.~\citep{2016suvorova}.
This suggests that the \bstat{} successfully exploits
all the orbital phase information to produce a nearly optimal outcome for a
semi-coherent algorithm, i.e. it analyses the orbital motion without any
degradation in sensitivity relative to an isolated source. The only information
it neglects is the phase continuity of the carrier wave at $f_0$ from one HMM
step to the next.
We quantify the optimality of the \bstat{} further in
Appendix~\ref{app:crlb}
via
an analytic calculation of the Cram\'er-Rao lower
bound.

\begin{table}
	\caption{Outcome of Viterbi tracking with
	the \bstat{} in synthetic data containing
	spin-wandering injections with the parameters in
	Table~\ref{tbl:synthparams}, $\Tobs =
	370\,\mathrm{d}$, $\Tdrift = 10\,\mathrm{d}$, and wave strain $h_0$.
	The root-mean-square error $\epsilon_{f_\star}$ between $f_\star(t)$ and
	the optimal path is quoted in columns 3 and 4.
	}
	\label{tbl:synthresults}
	\begin{tabular}{ccccc}
		\hline
		$h_0$ $(10^{-26})$ & Detect? & $\epsilon_{f_\star}$ (Hz) & $\epsilon_{f_\star} / \Delta\fdrift$ \\
		\hline
		8.0 & \checkmark & $3.54 \times 10^{-7}$ & $6.12 \times 10^{-1}$ \\
		5.0 & \checkmark & $3.55 \times 10^{-7}$ & $6.14 \times 10^{-1}$ \\
		4.0 & \checkmark & $3.73 \times 10^{-7}$ & $6.45 \times 10^{-1}$ \\
		2.0 & \checkmark & $5.80 \times 10^{-7}$ & $1.00 \times 10^{0\phantom{-}}$ \\
		1.5 & $\times$   & $1.91 \times 10^{-1}$ & $3.31 \times 10^{5\phantom{-}}$ \\
		\hline
	\end{tabular}
\end{table}

\subsection{Viterbi score}
\label{ssec:viterbiscore}
Once the HMM tracker finds the optimal path, it remains to decide if the
path constitutes a detection.
We define the Viterbi score $S$, such that the log likelihood of the optimal
path exceeds the mean log likelihood of all paths in the relevant sub-band (1
Hz, say, or whatever subdivision makes a broadband search practical) by $S$
standard
deviations, viz.
\begin{align}
	S = \frac{\ln \delta_{q^{\star}}(t_{N_T}) -
	\mu_{\ln\delta(t_{N_T})}}{\sigma_{\ln\delta(t_{N_T})}}
	\label{eqn:defS}
\end{align}
with
\begin{align}
	\mu_{\ln\delta(t_{N_T})} = N_Q^{-1} \sum_{i=1}^{N_Q} \ln\delta_{q_i}(t_{N_T})
\end{align}
and
\begin{align}
	\sigma_{\ln\delta(t_{N_T})}^2 = N_Q^{-1} \sum_{i=1}^{N_Q} [
		\ln\delta_{q_i}(t_{N_T}) - \mu_{\ln\delta(t_{N_T})} ]^2.
\end{align}
[The symbol $\delta$ is defined in equation~(\ref{eqn:vitdefdelta}).]
We then establish a threshold $\Sth$,
and claim a detection
for $S > \Sth$.
The threshold determines the false alarm
probability $\falsealarm$.
Selecting a desired false dismissal probability, $\falsedismissal$, then determines the
weakest signal we can reliably detect.

Appendix~\ref{app:far} discusses in detail the PDF of
the terminal Viterbi probabilities $\delta_{q_i}(t_{N_T})$
in~(\ref{eqn:vitdefdelta}). To the authors' knowledge, an analytic formula for
the PDF does not exist in the literature; the calculation is rendered difficult
by the correlations between Viterbi paths and the nonlinear maximisation step
in the algorithm. Appendix~\ref{app:far} presents an empirical fit to the
associated cumulative distribution function in the form of a Gumbel law,
motivated by asymptotic results from extreme value theory \citep{2004sornette}.
The two parameters of the fit (denoted by $a$ and $b$ in
Appendix~\ref{app:far}) are tabulated as functions of $N_Q$ and $N_T$ in
Table~\ref{tab:c:rmse} in the
appendix.

Table~\ref{tbl:scorefar}
presents $\falsealarm$ as a function of $\Sth$ for six convenient, representative,
integer
thresholds.
For the searches described in this
paper, we choose
$\Sth = 7$, which corresponds to
$\falsealarm = 7.1 \times 10^{-3}$, close to the false alarm probability of
$1\%$ per cent appearing commonly in the literature.
The choice $\Sth = 7$ also matches
what was done in Ref.~\citep{2016suvorova}, for ease of comparison.

\begin{table}
	\caption{False alarm probability $\falsealarm$ versus Viterbi score threshold
	$\Sth$ in equation~(\ref{eqn:defS}).}
	\label{tbl:scorefar}
	\begin{tabular}{lc} \hline
		$\Sth$ & $\falsealarm$ \\
		\hline
		5  & $9.4 \times 10^{-2}$ \\
		6  & $2.6 \times 10^{-2}$ \\
		7  & $7.1 \times 10^{-3}$ \\
		8  & $2.0 \times 10^{-3}$ \\
		9  & $5.0 \times 10^{-4}$ \\
		10 & $1.5 \times 10^{-4}$ \\
		\hline
	\end{tabular}
\end{table}

\subsection{Sensitivity to orbital parameters}
\label{ssec:parameters}
Electromagnetic observations of LMXBs play an important role in narrowing down
the range of possible values of the projected semimajor axis $a_0$ and
reference orbital phase $\phi_a$ \citep{2016premachandra,2014galloway}.
Typically, however, the range is wider than the resolution of the
\bstat{}--based HMM,
and a search over multiple templates within the range is still required.

\begin{figure*}
	\includegraphics[width=\textwidth]{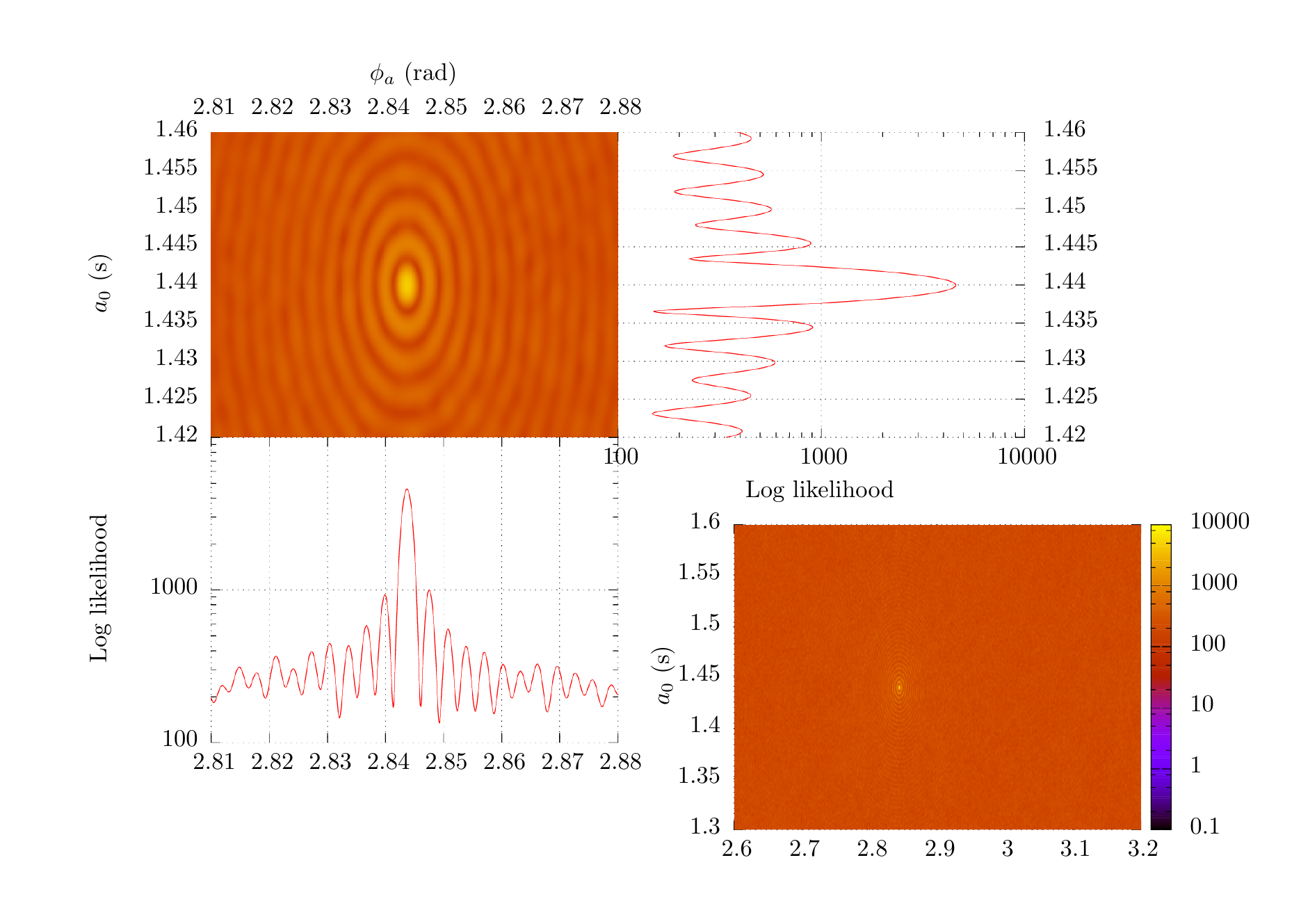}
	\caption{Log likelihood as a function of
	reference orbital phase
	$\phi_a$ and
	projected semi-major axis $a_0$
	for a search with $N_T = 37$ using two interferometers
	on synthetic data, for the parameters in Table~\ref{tbl:synthparams}
	with $h_0 = 8\times10^{-26}$.
	(Top left.) Contour plot of log likelihood on the $\phi_a$ - $a_0$ plane
	near the injection.
	The colour scale shows the log likelihood and is defined by the colour bar in the
	bottom right panel.
	(Top right.) Cross-section of log likelihood versus $a_0$ (units: s). The cross
	section is at $\phi_a = 0.6080\,\mathrm{rad}$, chosen to pass through the peak.
	(Bottom left.) Cross-section of log likelihood versus $\phi_a$ (units: rad).
	The cross section is at $a_0 = 1.3138\,\mathrm{s}$, again chosen to pass through the
	peak.
	(Bottom right.) Contour plot showing the entire parameter range searched.
	No spurious
	detections are found away from the injection parameters.
	}
	\label{fig:circle}
\end{figure*}

Figure~\ref{fig:circle} quantifies the resolution of the \bstat{}--based HMM in $a_0$ and
$\phi_a$ to help
fix the template spacing.
The figure is drawn for
a synthetic signal with $h_0 = 8\times 10^{-26}$, with parameters as in
Table~\ref{tbl:synthparams}.
It plots
log likelihood for
$1.3 \leq a_0/\mathrm{s} \leq 1.6$, $2.6 \leq \phi_a/\mathrm{rad} \leq
3.2$, and $N_T = 37$.
All other parameters,
including the search frequency $f_0 = f_\star$, are held fixed at their injected values.
The grid resolution is
$1\times10^{-4}\,\mathrm{s}$ for $a_0$
and
$1.5259 \times 10^{-5}\,\mathrm{rad}$ for $\phi_a$.
The figure shows a clear peak at the injected values of $a_0$ and
$\phi_a$. The peak is surrounded by rings (particularly visible in the zoomed
upper-left panel), which arise for two reasons: (i)
the decision in equation~(\ref{eqn:f2abstat})
to sum a finite number of sidebands, which introduces a $\sinc$-function
envelope as $a_0$ changes, and (ii) the finite observation time,
which introduces a $\sinc$-function envelope,
as $\phi_a$ changes.
There are
no false peaks away from the injected values.

The \bstat{} is sensitive to errors in $a_0$. For example,
a 10\% error in the measured value of $a_0$ causes a drop
of two orders of magnitude in the log likelihood, while
the \cstat{} sees
a 10\% reduction in detection probability for the same situation (see Figure~4 in
Ref.~\citep{2014sammut}).
The $a_0$ range covered in Figure~\ref{fig:circle}
is comparable to the uncertainty in the electromagnetic
measurement of $a_0$ at
the time Stage~I of the Sco~X-1 MDC was run \citep{2015messenger,2002steeghs}.
Since
then, the uncertainty has increased to
$0.36\,\mathrm{s}$--$3.25\,\mathrm{s}$ (Z. Wang et al., private communication). 
The computational cost scales linearly with the range of $a_0$.

\section{Sco X-1 mock data challenge: A ``realistic'' example}
\label{sec:mdc}
The next stage in validating the \bstat{}--based HMM is to engage in Stage~I
(version~6) of the Sco~X-1
MDC
\citep{2015messenger}. The MDC is based on a mock observational
dataset intended to simulate the noise level and duty cycle of Advanced LIGO.
Stage~I of the MDC
comprises 50 Sco~X-1--type signals without spin wandering injected into Gaussian noise.
\citep{2015messenger}
Stage~II of the MDC is currently being prepared and is planned to include
spin wandering.
The
MDC
establishes
a standard to compare the \bstat{} HMM tracker against the
Bessel-weighted \fstat{} \cite{2016suvorova},
CrossCorr \citep{2008dhurandhar,2011chung,2015whelan,2016sun},
TwoSpect \citep{2014lsctwospect,2011goetz,2016meadors},
Radiometer \citep{2011lscradiometer},
Sideband \citep{2015lscyoungsnr,2014sammut,2007messenger} and
Polynomial \citep{2010putten} pipelines.

The parameters of the 50 Stage~I MDC injections are listed in
Table~III of Ref.~\citep{2015messenger}.
Originally, these 50 injections were ``closed'', i.e. their parameters were kept
secret to
enable a blind comparison. The TwoSpect, Radiometer, Sideband and Polynomial
pipelines performed the test under closed conditions as reported in
Ref.~\citep{2015messenger}, while CrossCorr and the
Bessel-weighted \fstat{} \citep{2016suvorova} participated after the release of
the parameters
under self-blinded conditions, as we propose to do here. Participants
in the original tests were asked to assume, that the injections experience
spin wandering (although they do not), with the Sideband search being
restricted to $\Tobs = 10\,\mathrm{d}$ as a result \citep{2015messenger}.
We use the transition
matrix~(\ref{eqn:transmat}) to replicate this mode of operation.

The orbital period $P$ of Sco~X-1 is measured to $\pm 0.04\,\mathrm{s}$
\citep{2014galloway}.
The error in sideband frequency, for all $M$ sidebands, must be less
than one \bstat{} frequency bin, which
limits the
allowed uncertainty in $P$ to $|\Delta P| \leq P^2 /
(M\Tdrift) \leq 0.2\,\mathrm{s}$.\footnote{This formula is the same as
equation~(58) in Ref.~\citep{2014sammut} for $\Delta \fdrift =
1/(2\Tdrift)$.}
This suggests that searching over $P$ is unnecessary. For the search
described in this section, we assume $P = 68023.70\,\mathrm{s}$ for all
injections \citep{2014galloway} (cf. Table~II in Ref.~\citep{2015messenger}).

We divide the year-long dataset, starting at GPS time $1\,230\,338\,490$,
into $N_T = 37$ blocks with $\Tdrift = 10\,\mathrm{d}$.
Data
from two simulated interferometers (H1 and L1) are used in the analysis below.

\subsection{Single block: $N_T = 1$}
\label{ssec:mdcsingleblock}
The first step is to ask how many injections are detected using the first block
only ($N_T = 1$).
We find that the answer is 43 out of 50.
The exceptions are those with index 41, 48, 57, 64, 72, 73 and 90. By way of
comparison, the Bessel-weighted \fstat{} with $N_T = 1$ and two interferometers
detects only 12 signals \citep{2016suvorova}, and the \cstat{} with $N_T = 1$
and three interferometers detects 16 signals \citep{2015messenger}.

Detailed test results are presented in Table~\ref{tbl:mdclist}.
The table lists the
injection
parameters ($f_\star$, $a_0$, $\phi_a$, $h_0$)
as well as the absolute (as opposed to relative)
errors $\epsilon_{f_\star}$, $\epsilon_{a_0}$ and
$\epsilon_{\phi_a}$ in the recovered values for $f_\star$, $a_0$ and
$\phi_a$ respectively. For $a_0$ and $\phi_a$, the search returns single
grid values, so $\epsilon_{a_0}$ and $\epsilon_{\phi_a}$ are defined as the
signed difference between the injected and recovered values. For $f_\star$,
which wanders in general,
we
define
$\epsilon_{f_\star}$ as the root-mean-square error between the injected and
optimal paths for however
many blocks are needed to achieve a detection, in preparation for the analysis
in Section~\ref{ssec:mdcmultiblock} with $N_T > 1$.

For most signals, the error in $a_0$ and $\phi_a$ is smaller than the
bin size for those parameters ($1 \times 10^{-4}\,\mathrm{s}$ for $a_0$ and $1.5259
\times 10^{-5}\,\mathrm{rad}$
for $\phi_a$),
so the error is the difference between the bin boundary
and the injection parameter.
The log likelihood
peaks sharply, as the estimate of $\phi_a$ improves, so it may be
possible to
improve the sensitivity somewhat
by estimating $\phi_a$ more precisely.
We defer to future work the task of determining the optimal template spacing
for a given mismatch using the parameter space metrics derived in
Ref.~\citep{2015leaci}.
For 13 injections, the RMS error between the optimal and injected paths is less
than the \bstat{} frequency bin width,
$\Delta \fdrift = 5.8 \times 10^{-7}\,\mathrm{Hz}$, and seven more
have $1.0\Delta\fdrift \leq \epsilon_{f_star} \leq 1.5 \Delta \fdrift$.
The remaining 23 injections are detected with frequency error $19\Delta
\fdrift < \epsilon_{f_\star} < 34 \Delta \fdrift$.
The distance (in frequency space) between this group and the group with error
less than $1.5\Delta \fdrift$ corresponds roughly to the frequency separation
between sidebands.

\begin{figure}
	\includegraphics[width=0.49\textwidth]{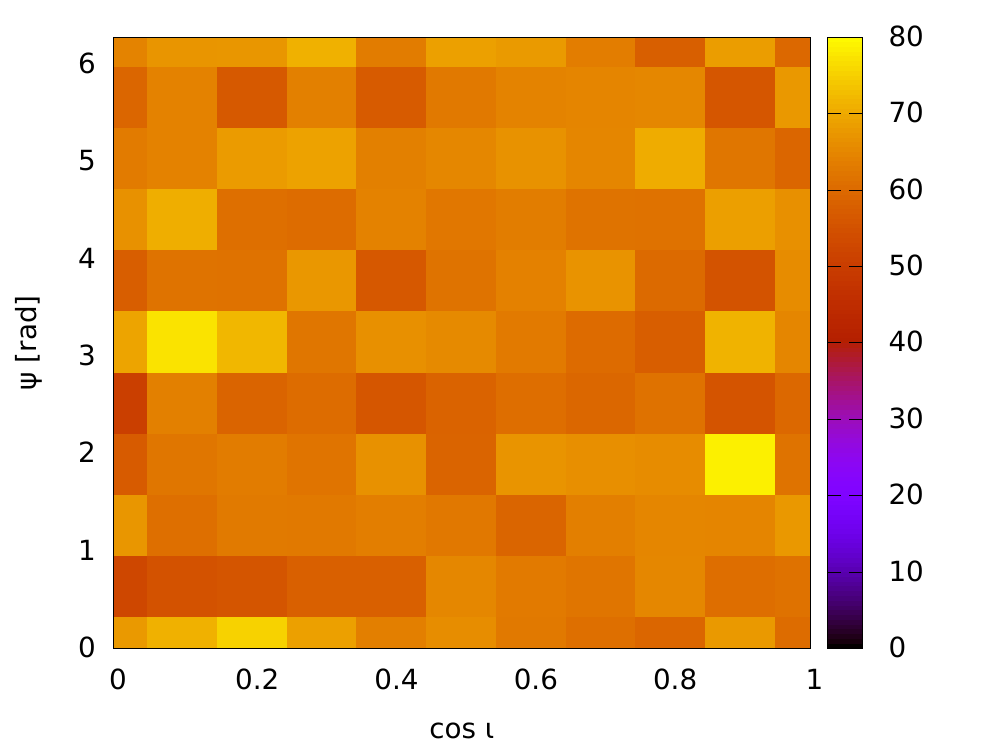}
	\caption{Log likelihood (for $N_T = 1$)
	versus orientation angles $\cos\iota$ and $\psi$, holding
	$\heff = 8\times10^{-26}$ fixed and
	all other parameters as in Table~\ref{tbl:synthparams}.
	Each grid cell is an average over $10^2$ noise realisations.}
	\label{fig:psicosi}
\end{figure}

The characteristic wave strain
$h_0$
influences
detectability,
in conjunction with
the source inclination angle $\iota$,
which enters the plus and cross polarisations differently.
A popular, approximate proxy
for signal strength, given by equation~(19) in Ref.~\citep{2015messenger}, is
the effective characteristic wave strain
\begin{align}
	\heff = h_0 2^{-1/2} \{[(1+ \cos^2 \iota)/2]^2 + \cos^2\iota\}^{1/2}.
\end{align}
To test if $\heff$ captures faithfully the joint dependence of detectability on
$h_0$ and $\iota$,
we generate synthetic
signals for $0 \leq \cos\iota \leq 1$ and $0 \leq \psi \leq 2\pi$,
while holding $\heff$ constant (that is, with different values of $h_0$ for
each $\iota$ value) and calculate the \bstat{}.
If $\heff$ is a perfect proxy, we expect
the same
\bstat{} output for all $\iota$.
The
results of 100
Monte-Carlo
realisations per $(\cos\iota, \psi)$ pair are plotted
in Figure~\ref{fig:psicosi}. Indeed the \bstat{} score is roughly
constant, showing no discernible pattern across the full parameter space,
and fluctuating by $\leq 12\%$.

\begin{figure}
	\begin{subfigure}[b]{0.49\textwidth}
		\includegraphics[width=\textwidth]{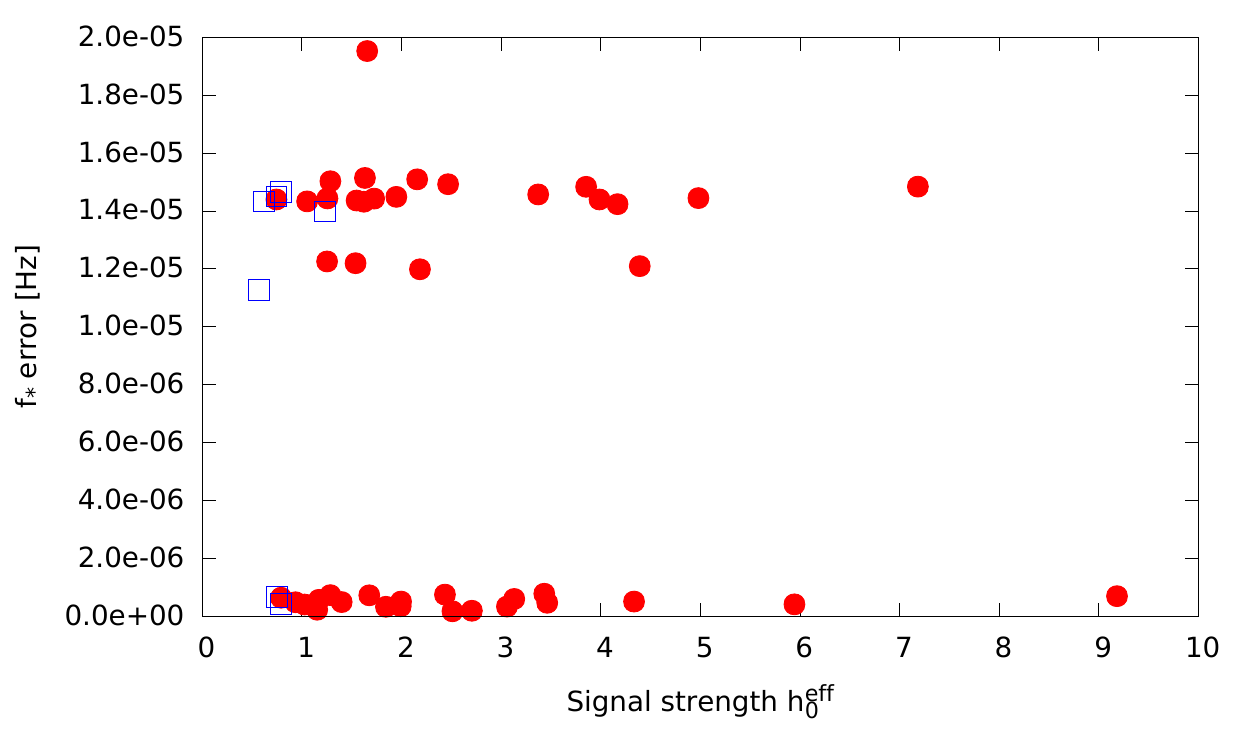}
		\caption{}
		\label{sfig:5:freq}
	\end{subfigure}
	\hfill
	\begin{subfigure}[b]{0.49\textwidth}
		\includegraphics[width=\textwidth]{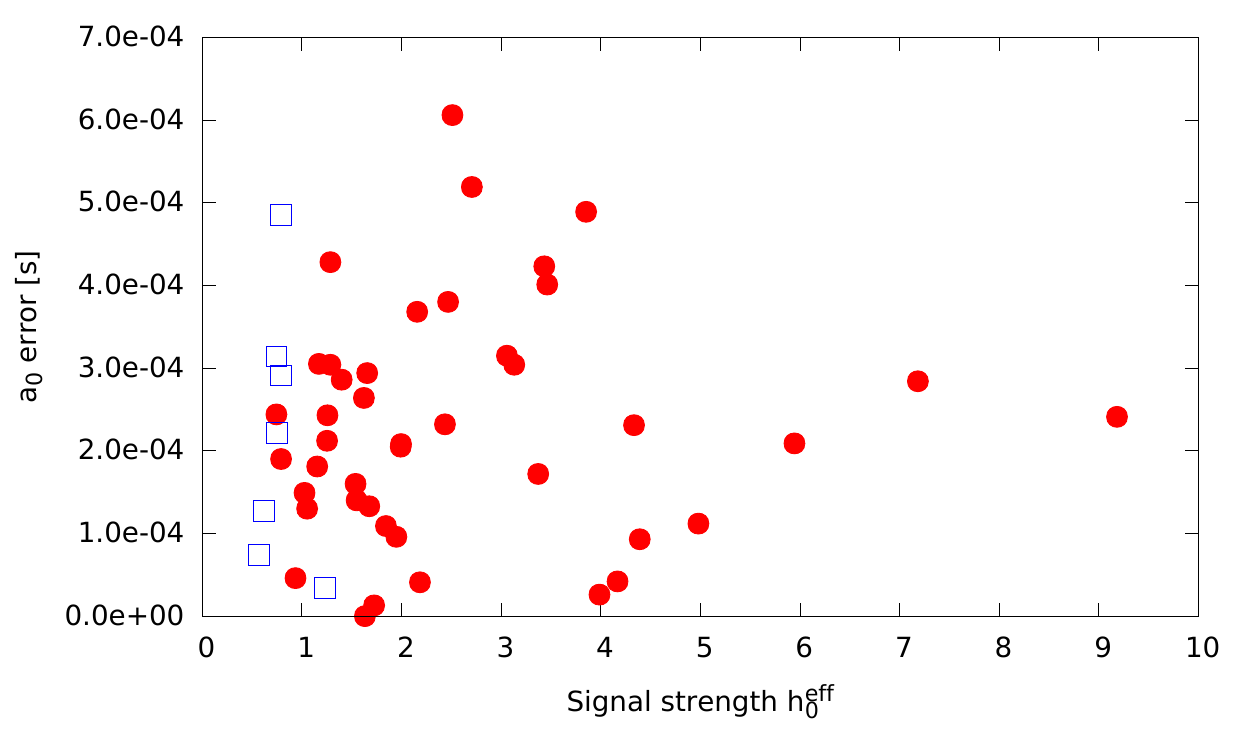}
		\caption{}
		\label{sfig:5:asini}
	\end{subfigure}

	\begin{subfigure}[b]{0.49\textwidth}
		\includegraphics[width=\textwidth]{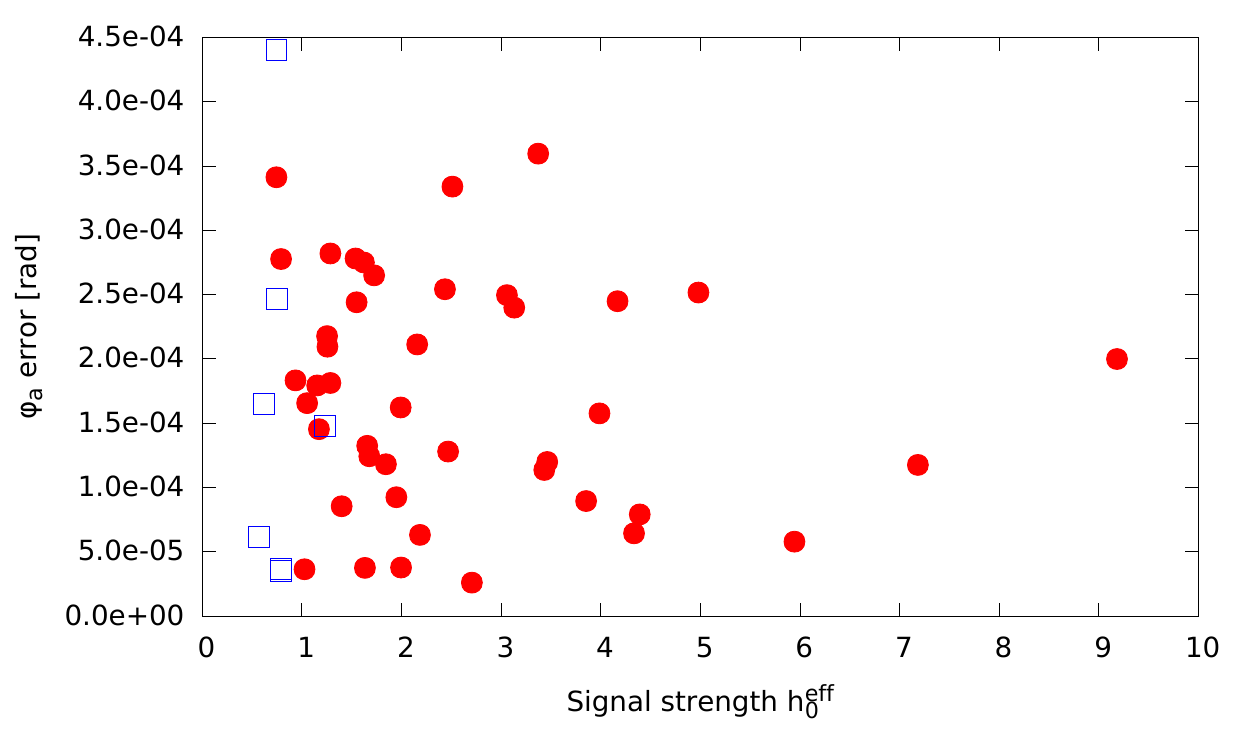}
		\caption{}
		\label{sfig:5:phi}
	\end{subfigure}
	\caption{Accuracy of the \bstat{}--based HMM applied to Stage~I of the Sco~X-1
	MDC versus effective characteristic wave strain $\heff$.
	(a) Root-mean-square error in $f_\star$ (Hz).
	(b) Absolute unsigned error in $a_0$ (in s).
	(c) Absolute unsigned error in $\phi_a$ (in rad).
	The cluster in (a) at $\epsilon_{f_\star} \approx
	1.4\times10^{-5}\,\mathrm{Hz}$ corresponds to the orbital sideband
	separation $P^{-1}$.
	All 50 MDC injections are shown: 43 are detected with $N_T = 1$ (filled
	red circles; see section~\ref{ssec:mdcsingleblock}), while seven require $1
	< N_T \leq 37$ (open blue squares; see section~\ref{ssec:mdcmultiblock}).
	}
	\label{fig:mdcerr}
\end{figure}

Figure~\ref{fig:mdcerr} summarises
the error estimates in
Table~\ref{tbl:mdclist}. It displays $\epsilon_{f_\star}$, $\epsilon_{a_0}$ and
$\epsilon_{\phi_a}$
plotted against $\heff$.
The vertical ``step'' in $\epsilon_{f_\star}$ visible in
Figure~\ref{sfig:5:freq} corresponds
to the
sideband separation $P^{-1}$.
In Figures~\ref{sfig:5:asini}
and~\ref{sfig:5:phi}, $\epsilon_{a_0}$ and $\epsilon_{\phi_a}$
are comparable to the grid resolution, with all 50 injections having errors less
than
seven ($\epsilon_{a_0}$) or five ($\epsilon_{\phi_a}$) search bins.
There is no apparent correlation between the errors and $\heff$ amongst the
injections that are detected successfully.

\subsection{Multiple blocks, $1 < N_T \leq 37$}
\label{ssec:mdcmultiblock}
For the seven out of 50 injections that are not detected with $N_T = 1$,
we do HMM tracking for $\Tdrift = 10\,\mathrm{d}$
and $1 < N_T \leq 37$. The search parameters and prior are the same as in
Section~\ref{ssec:mdcsingleblock}. We successfully detect all seven remaining
injections, with $N_T =
3$ for injection 73, $N_T = 13$ for injection $90$, and the others in between,
as in Table~\ref{tbl:mdcspans}.

\begin{figure}
	\includegraphics[width=0.49\textwidth]{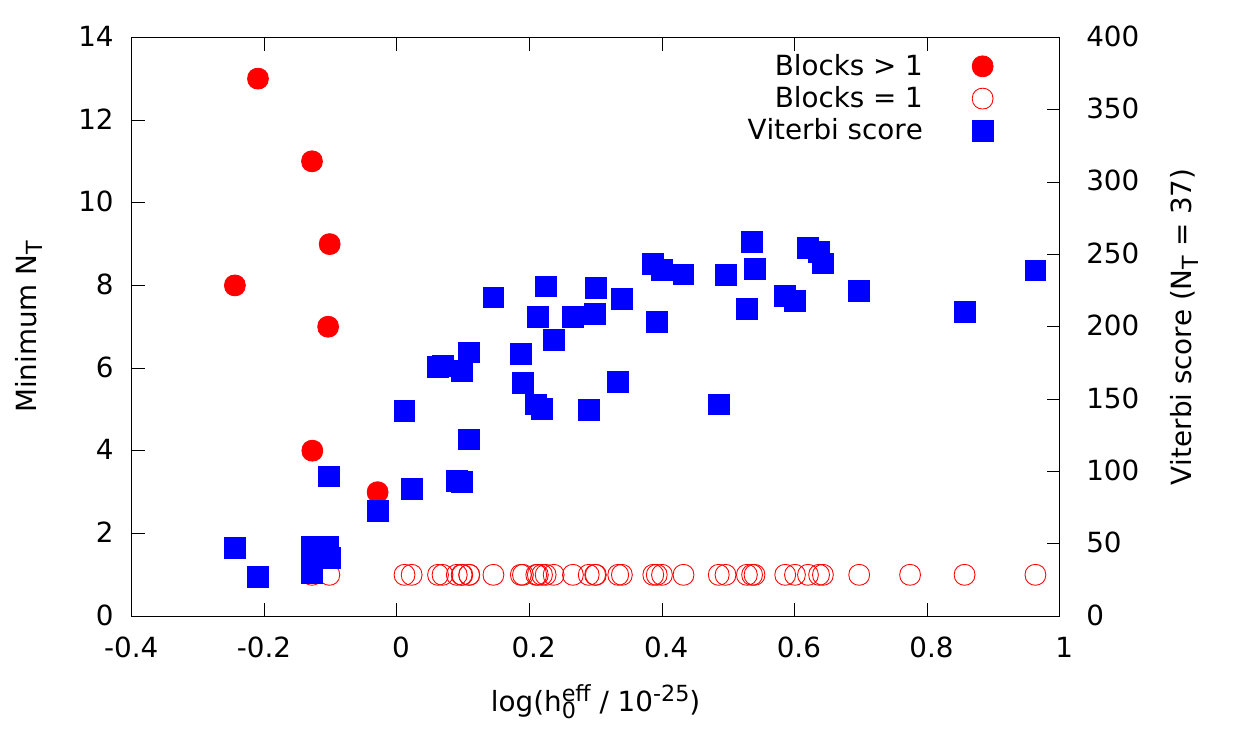}
	\caption{Minimum number of 10-d data blocks $N_T$ for a detection (left axis; red
	circles, filled for $N_T > 1$, open for $N_T = 1$) and Viterbi score $S$
	with $N_T = 37$ (right axis; blue squares) versus the logarithm of the
	effective characteristic wave strain for 50 injections in Stage~1 of the
	Sco~X-1 MDC.
	}
	\label{fig:heffspan}
\end{figure}

Figure~\ref{fig:heffspan} summarises the results in Tables~\ref{tbl:mdclist}
and~\ref{tbl:mdcspans}. It plots all 50 injections twice against
$\heff$. The red circles
(left axis)
indicate the minimum $N_T$ required
to achieve a detection.
Most
injections are
detected
in a single block (open red circles). The seven injections requiring multiple
blocks (filled red circles) are
all weak signals, along the left-hand border of the plot. The blue
squares (right axis) show
the Viterbi score $S$ after processing the full year of data, even if the
signal is detected with $N_T < 37$. There is
a positive correlation between $S$ and $\heff$,
with $S \propto \heff$ roughly for $S \lesssim 2.6\times10^2$.
The Viterbi score for injection 66 ($S = 522$) lies outside the range of the
graph, due to a lucky coincidence of a relatively strong signal and a $\phi_a$
value that happens to lie close to the search grid.

\begin{table*}
	\vspace{-0.8cm}
	\caption{Results of \bstat{} HMM tracking of the 50 closed signals in
	Stage~I (version~6) of the
	Sco~X-1 MDC, ordered by ascending injection frequency $f_\star$.
	The index is copied from Ref.~\citep{2015messenger}.
	Signal strength is quantified by $h_0$ (traditional gravitational wave
	strain in Ref.~\citep{1998jks,2015messenger}) and $\heff$
	(Ref.~\citep{2015messenger}).
	The recovered orbital parameters $a_0$ and $\phi_a$ and their signed,
	absolute errors appear in columns six to nine.
	}
	\label{tbl:mdclist}
	\begin{tabular}{ccccccccc}
	\hline \hline
		Index & $h_0$            & $h^{\mathrm{eff}}_0$ & $f_\star$ & $\epsilon_{f_\star}$ & $a_0$ & $\epsilon_{a_0}$ & $\phi_a$  & $\epsilon_{\phi_a}$ \\
		      & ($    10^{-25}$) & ($10^{-25}$)         & (Hz)      &           (Hz)       & (s)   &       (s)        & (rad)     &          (rad)      \\
	\hline
1 & 4.160101 & 2.706 & 54.498391348174 & 1.853E-07 & 1.379519 & 5.190E-04 & 0.564832303 & -2.606E-05 \\
2 & 4.044048 & 2.511 & 64.411966012332 & 1.681E-07 & 1.764606 & 6.060E-04 & 0.572064312 & -3.339E-04 \\
3 & 3.565197 & 3.463 & 73.795580913582 & 4.603E-07 & 1.534599 & -4.010E-04 & 0.585084391 & 1.198E-04 \\
5 & 1.250212 & 1.154 & 93.909518008164 & 2.244E-07 & 1.520181 & 1.810E-04 & 0.633165725 & -1.794E-04 \\
11 & 3.089380 & 1.399 & 154.916883586097 & 4.866E-07 & 1.392286 & 2.860E-04 & 0.576082666 & -8.545E-05 \\
14 & 2.044140 & 1.286 & 183.974917468730 & 7.248E-07 & 1.509696 & -3.040E-04 & 0.577142828 & -2.819E-04 \\
15 & 11.763777 & 4.169 & 191.580343388804 & 1.424E-05 & 1.518142 & 4.200E-05 & 0.599799259 & -2.449E-04 \\
17 & 3.473418 & 1.253 & 213.232194220000 & 1.225E-05 & 1.310212 & 2.120E-04 & 0.578899085 & 2.177E-04 \\
19 & 6.030529 & 2.437 & 233.432565653291 & 7.465E-07 & 1.231232 & 2.320E-04 & 0.596020206 & -2.541E-04 \\
20 & 9.709634 & 3.434 & 244.534697522529 & 7.748E-07 & 1.284423 & 4.230E-04 & 0.617523371 & -1.137E-04 \\
21 & 1.815111 & 0.792 & 254.415047846878 & 6.374E-07 & 1.072190 & 1.900E-04 & 0.595996707 & -2.776E-04 \\
23 & 2.968392 & 1.677 & 271.739907539784 & 7.173E-07 & 1.442867 & -1.330E-04 & 0.598663241 & -1.243E-04 \\
26 & 1.419173 & 1.172 & 300.590450155009 & 5.630E-07 & 1.258695 & -3.050E-04 & 0.610242598 & 1.453E-04 \\
29 & 4.274554 & 3.131 & 330.590357652653 & 5.968E-07 & 1.330696 & -3.040E-04 & 0.580326474 & -2.398E-04 \\
32 & 10.037770 & 4.391 & 362.990820993568 & 1.209E-05 & 1.611093 & 9.300E-05 & 0.573105599 & 7.911E-05 \\
35 & 16.401523 & 9.183 & 394.685589797695 & 6.921E-07 & 1.313759 & -2.410E-04 & 0.608012394 & -1.999E-04 \\
36 & 3.864262 & 1.539 & 402.721233789014 & 1.219E-05 & 1.254840 & -1.600E-04 & 0.602207114 & 2.780E-04 \\
41 & 1.562041 & 0.746 & 454.865249156175 & 6.744E-07 & 1.465778 & -2.220E-04 & 0.605945666 & 2.466E-04 \\
44 & 2.237079 & 1.996 & 483.519617972096 & 5.065E-07 & 1.552208 & 2.080E-04 & 0.590657162 & 3.774E-05 \\
47 & 4.883365 & 1.992 & 514.568399601819 & 3.425E-07 & 1.140205 & 2.050E-04 & 0.563763897 & 1.622E-04 \\
48 & 1.813016 & 0.745 & 520.177348201609 & 1.451E-05 & 1.336686 & -3.140E-04 & 0.563161604 & -4.401E-04 \\
50 & 1.092771 & 1.027 & 542.952477491471 & 4.038E-07 & 1.119149 & 1.490E-04 & 0.542275328 & 3.644E-05 \\
51 & 9.146386 & 3.372 & 552.120598886904 & 1.457E-05 & 1.327828 & -1.720E-04 & 0.573295251 & -3.596E-04 \\
52 & 2.785731 & 1.550 & 560.755048768919 & 1.436E-05 & 1.792140 & 1.400E-04 & 0.594773666 & -2.440E-04 \\
54 & 1.517530 & 1.256 & 593.663030872532 & 1.443E-05 & 1.612757 & -2.430E-04 & 0.569675332 & -2.095E-04 \\
57 & 1.576918 & 0.788 & 622.605388362863 & 4.347E-07 & 1.513291 & 2.910E-04 & 0.608877237 & 3.658E-05 \\
58 & 3.416297 & 1.287 & 641.491604906276 & 1.503E-05 & 1.584428 & 4.280E-04 & 0.602738791 & 1.813E-04 \\
59 & 8.834794 & 4.981 & 650.344230698489 & 1.444E-05 & 1.677112 & 1.120E-04 & 0.550155435 & -2.516E-04 \\
60 & 2.960648 & 2.467 & 664.611446618250 & 1.492E-05 & 1.582620 & -3.800E-04 & 0.568756259 & 1.280E-04 \\
61 & 6.064238 & 2.158 & 674.711567789201 & 1.509E-05 & 1.499368 & 3.680E-04 & 0.626850596 & -2.113E-04 \\
62 & 10.737497 & 3.853 & 683.436210983289 & 1.483E-05 & 1.269511 & -4.890E-04 & 0.585682431 & 8.954E-05 \\
63 & 1.119028 & 0.745 & 690.534687981171 & 1.440E-05 & 1.518244 & 2.440E-04 & 0.587764962 & -3.412E-04 \\
64 & 1.599528 & 0.570 & 700.866836291234 & 1.129E-05 & 1.399926 & -7.400E-05 & 0.571080095 & -6.145E-05 \\
65 & 8.473643 & 4.334 & 713.378001688688 & 5.023E-07 & 1.145769 & -2.310E-04 & 3.981714377 & 6.434E-05 \\
66 & 9.312048 & 5.944 & 731.006818153273 & 4.061E-07 & 1.321791 & -2.090E-04 & 3.937174208 & -5.789E-05 \\
67 & 4.579697 & 1.623 & 744.255707971300 & 1.432E-05 & 1.677736 & -2.640E-04 & 0.619168642 & 2.749E-04 \\
68 & 3.695848 & 1.844 & 754.435956775916 & 3.240E-07 & 1.413891 & -1.090E-04 & 0.577934937 & -1.181E-04 \\
69 & 2.889282 & 1.053 & 761.538797037770 & 1.433E-05 & 1.626130 & 1.300E-04 & 0.642604270 & -1.656E-04 \\
71 & 2.922576 & 1.232 & 804.231717847467 & 1.398E-05 & 1.652034 & 3.400E-05 & 0.614347724 & -1.478E-04 \\
72 & 1.248093 & 0.792 & 812.280741438401 & 1.466E-05 & 1.196485 & 4.850E-04 & 0.612575356 & -3.521E-05 \\
73 & 2.443983 & 0.936 & 824.988633484129 & 4.802E-07 & 1.417154 & -4.600E-05 & 0.545563765 & 1.833E-04 \\
75 & 7.678400 & 3.987 & 862.398935287248 & 1.440E-05 & 1.567026 & 2.600E-05 & 3.958458316 & 1.576E-04 \\
76 & 3.260143 & 1.725 & 882.747979842807 & 1.443E-05 & 1.462487 & -1.300E-05 & 0.648061399 & 2.650E-04 \\
79 & 4.680848 & 1.656 & 931.006000308958 & 1.953E-05 & 1.491706 & -2.940E-04 & 0.598919953 & 1.324E-04 \\
83 & 5.924668 & 2.186 & 1081.398956458276 & 1.198E-05 & 1.198541 & 4.100E-05 & 0.598724345 & -6.321E-05 \\
84 & 11.608892 & 7.184 & 1100.906018344283 & 1.484E-05 & 1.589716 & -2.840E-04 & 0.609351448 & -1.176E-04 \\
85 & 4.552730 & 1.633 & 1111.576831848269 & 1.514E-05 & 1.344790 & 0.000E+00 & 0.623329562 & 3.758E-05 \\
90 & 0.684002 & 0.618 & 1193.191890630547 & 1.433E-05 & 1.575127 & 1.270E-04 & 0.636321462 & -1.652E-04 \\
95 & 4.293322 & 3.059 & 1324.567365220908 & 3.271E-07 & 1.591685 & -3.150E-04 & 0.587727432 & 2.496E-04 \\
98 & 5.404060 & 1.948 & 1372.042154535880 & 1.449E-05 & 1.315096 & 9.600E-05 & 0.640164126 & -9.243E-05 \\
\hline
	\end{tabular}
\end{table*}

\begin{table}
	\caption{Minimum number of data blocks $N_T$ required to detect the seven MDC
	injections that cannot be detected with $N_T = 1$. Indices refer to
	Table~\ref{tbl:mdclist}.
	A detection is claimed when the Viterbi score exceeds the mean by at least
	seven standard deviations ($S > 7$).}
	\label{tbl:mdcspans}
	\begin{tabular}{cr}
		\hline
		Index & Blocks \\
		\hline
41 & 4 \\
48 & 11 \\
57 & 7 \\
64 & 8 \\
72 & 9 \\
73 & 3 \\
90 & 13 \\
\hline
	\end{tabular}
\end{table}

\section{Conclusion}
In this paper, we
extend the HMM scheme for
tracking continuous-wave gravitational
radiation
from a neutron star undergoing spin wandering in an LMXB
described in
Ref.~\citep{2016suvorova}.
The new scheme tracks the orbital phase of the source by using a
frequency-domain matched filter, termed the \bstat{}, to compute the emission
probabilities at each HMM step. The \bstat{} sums the \fstat{} power in orbital
sidebands coherently by weighting each sideband by a suitable Bessel amplitude
and Fourier phase.
Monte-Carlo simulations in Gaussian noise with $S_h(f_\star)^{1/2} =
4\times10^{-24}\,\mathrm{Hz}^{-1/2}$
show that the \bstat{} HMM successfully detects
spin-wandering injections with wave strain $h_0 \gtrsim 2 \times
10^{-26}$ with two interferometers.
This equals the sensitivity achieved in Ref.~\citep{2016suvorova}
for isolated neutron stars; the \bstat{} succeeds in marshalling all the signal
power in orbital sidebands into a single frequency bin with essentially zero
leakage. Even better sensitivity will be achieved when combining three
interferometers.

When competing in self-blinded mode
in Stage~I of the Sco~X-1 MDC, the \bstat{} HMM detects all
50 signals, 43 of them using a single HMM step ($10\,\mathrm{d}$ of data).
It estimates $f_\star$, $a_0$ and $\phi_a$ to accuracies of
$\epsilon_{f_\star} < 2 \times 10^{-5}\,\mathrm{Hz}$,
$\epsilon_{a_0} < 6 \times 10^{-5}\,\mathrm{s}$,
$\epsilon_{\phi_a} < 4 \times 10^{-4}\,\mathrm{rad}$
respectively. By
comparison, the CrossCorr, Bessel-weighted \fstat{} HMM, TwoSpect, Radiometer, Sideband and Polynomial
methods found 50, 41, 34, 28, 16 and 7 out of 50 signals, respectively.
The accuracy of $f_\star$ estimation by the
\bstat{} HMM is
roughly as good as the most accurate existing algorithms (CrossCorr,
Bessel-weighted HMM, TwoSpect).
The same is true for its accuracy of
$a_0$ estimation. (Radiometer and Polynomial do not estimate $a_0$.) A
comparison of the performance metrics for the seven algorithms listed above is
presented in
Table~\ref{tbl:algocomp}.
The \bstat{} HMM is the only scheme to be tested formally on
spin-wandering data, as reported here, although this will change when Stage~II
of the Sco~X-1 MDC is completed. We emphasise that several of the algorithms in
Table~\ref{tbl:algocomp} have undergone substantial refinement, since
Ref.~\citep{2015messenger} was published, e.g. the tuned TwoSpect method
\citep{2016meadors}.
When performance data are published for the refined algorithms,
some of the entries in Table~\ref{tbl:algocomp}
will require updating.

Although this paper focuses on spin wandering in LMXBs, the same methods are
likely to prove helpful when searching for isolated, nonaccreting neutron stars
as well. Radio timing experiments reveal that spin wandering is endemic in
rotation-powered pulsars, where it goes by the name of `timing noise'
\citep{2010hobbs,2010shannon}. Timing noise exhibits a red spectrum and is
autocorrelated on time-scales ranging from days to years
\citep{1980cordes,2012price}. Its physical origin is still debated, but it is
generically attributed to fluctuations in the structure of the magnetosphere
and/or superfluid interior
\citep{1986alpar,1987cheng,1990jones,2010hobbs,2010lyne,2014melatos}.
Until now, continuous-wave searches have handled timing noise in various ways.
All-sky searches for periodic signals from isolated neutron stars in the LIGO
(S5 and S6) and Virgo (VSR1 to VSR4) data sets, using various algorithms (e.g.,
loosely coherent, Hough, \fstat{}), typically consider a range of spin-down
rates \citep{2013lsceathome,2016lscallsky,2014lsctwospect,2016walsh}.
For these experiments, spin wandering effectively limits the maximum
observation time, before the phase model loses coherence with the source. The
same applies to directed \fstat{} and hierarchical searches pointed at young
supernova remnants and the Galactic centre respectively
\citep{2013lscgalcentre,2016lscorion,2010lsccasa}.
Coherent narrowband searches for objects like the Crab and Vela pulsars are
guided by radio pulsar timing ephemerides, so in principle the timing noise
is tracked electromagnetically
\citep{2016lscallsky,2008lsc,2014lscsummary,2013lsceathome,2011lscvela,2010lscknownpulsar}.
Even so there is no guarantee that the gravitational-wave-emitting quadrupole
is locked to the stellar crust and magnetic field and hence the radio emission;
a lag may exist between the two components and it may fluctuate stochastically
\citep{2013warszawski,2014melatos,2015melatosdouglass}. Coherent narrowband
searches usually safeguard against this eventuality by scanning a band of
frequency centred on the radio ephemeris, typically $\sim \pm
10^{-2}\,\mathrm{Hz}$ wide, without explicitly testing all possible frequency
wandering paths within the band, e.g., \citep{2015lsccrabvela}.
\citet{2015ashton} quantified the loss of sensitivity caused by timing noise in
ephemeris-guided narrowband searches, calculating the template mismatch as a
function of the total observation time. We will investigate the performance of
HMM frequency tracking in these contexts in future work.

\begin{table*}
	\caption{Comparison between the \bstat{} HMM (Viterbi~2.0) and other
	algorithms that participated in Stage~I of the Sco~X-1
	MDC. Viterbi~1.0 refers to the Bessel-weighted \fstat{} combined with the
	Viterbi HMM solver in Ref.~\citep{2016suvorova}. The table does not include
	unpublished performance improvements to several algorithms listed.}
	\label{tbl:algocomp}
	\begin{tabular}{lrrrrrrr}
		\hline
		         & Viterbi 2.0 & CrossCorr & Viterbi 1.0  & TwoSpect  & Radiometer & Sideband  & Polynomial \\
		\hline
		Hit rate (out of 50)
				 & 50        & 50        & 41        & 34        & 28         & 16        & 7 \\
		Best $h_0$ $(10^{-25})$
				 & 0.684     & 0.684     & 1.093     & 1.250     & 2.237      & 3.565     & 7.678 \\
		Best $h_0/\sqrt{f_\star}$ ($10^{-25}\,\mathrm{Hz}^{-1/2}$)
				 & 0.020     & 0.020     & 0.047     & 0.082     & 0.102      & 0.235     & 0.261 \\
		Typical $\epsilon_{f_\star}$ (Hz)
				 & $10^{-5}$ & $10^{-5}$ & $10^{-5}$ & $10^{-4}$ & $10^{-1}$  & $10^{-2}$ & $10^{-2}$ \\
		Typical $\epsilon_{a_0}$ (s)
		         & $10^{-4}$ & $10^{-4}$ & $10^{-4}$ & $10^{-2}$ & ---        & ---       & --- \\
		Typical run time (CPU-hr)
		         & $10^5$    & $10^6$    & $10^3$    & $10^5$    & $10^3$     & $10^3$    & $10^8$ \\
				 \hline
	\end{tabular}
\end{table*}

\section{Acknowledgements}
We thank the LIGO Scientific Collaboration Continuous Wave
Working Group for informative discussions.
The synthetic data for Stage~I of the Sco~X-1 MDC were prepared primarily by
Chris Messenger with the assistance of members of the MDC team.
\citep{2015messenger}
P.~Clearwater is supported by a Melbourne Research Scholarship and a CSIRO
Office of the Chief Executive
Postgraduate PhD Scholarship in Zettabyte Data Management.
L.~Sun is supported by an Australian Postgraduate Award.
This work was
supported by the Multi-modal
Australian ScienceS Imaging and Visualisation Environment (MASSIVE),
by Australian Research Council (ARC) Discovery
Project DP110103347, ARC Centre of Excellence CE170100004 and by the U.S.~Air
Force Office of Scientific Research under Grant No.~FA9550-12-1-0418.

\appendix

\begin{figure*}
	\begin{subfigure}[b]{0.49\textwidth}
		\includegraphics[width=\textwidth]{fig_1_c.pdf}
		\caption{}
		\label{fig:A1:besselwt}
	\end{subfigure}
	\hfill
	\begin{subfigure}[b]{0.49\textwidth}
		\includegraphics[width=\textwidth]{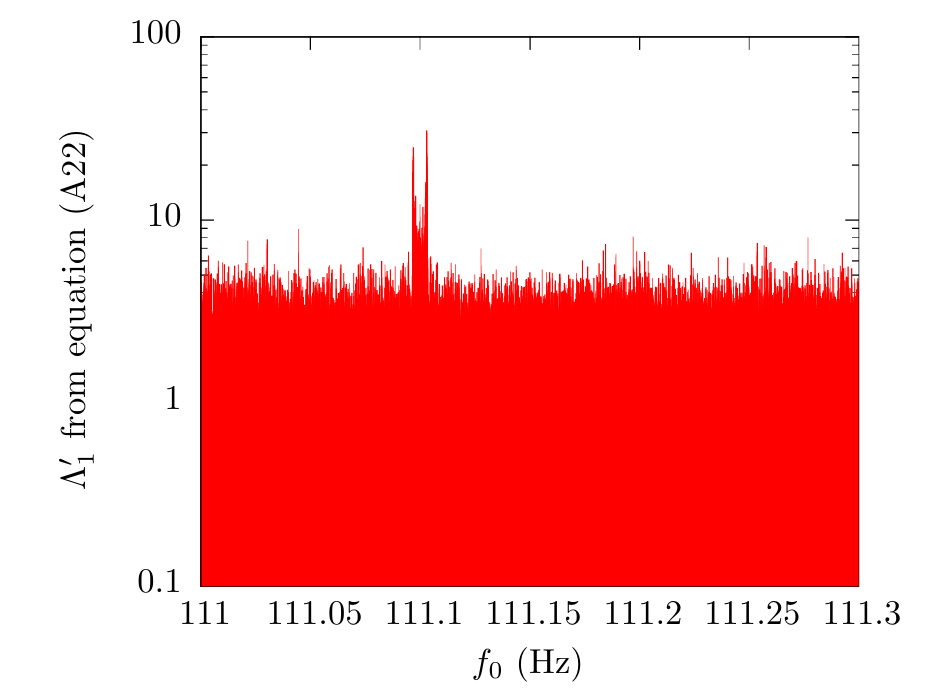}
		\caption{}
		\label{fig:A1:fppstat}
	\end{subfigure}
	\caption{Signature of an injected binary signal in~(a) the
	Bessel-weighted \fstat{}, described in Ref.~\citep{2016suvorova}
	and~(b) the quasi-maximum likelihood estimator,
	defined by~(\ref{eqn:a:lambda1simpl}). As in
	Figure~\ref{fig:comparison}, the plots are normalised such that the mean of the
	noise is unity. Parameters: as in
	Table~\ref{tbl:synthparams}, with $h_0 = 8 \times 10^{-25}$.
	}
\label{fig:a:fppstat}
\end{figure*}

\section{Quasi--maximum-likelihood generalisations of the \fstat{} for binary
sources}
\label{sec:app:qmle}

A true maximum-likelihood, frequency-domain estimator that generalises the
\fstat{} to handle binary sources has not yet been derived in the literature.
Such an estimator would maximise $\Lambda'_1$ in~(\ref{eqn:dfnlambda}) for the signal
model~(\ref{eqn:ht})--(\ref{eqn:h24}) and phase model
in~(\ref{eqn:phasemodel}) by solving for the optimal values of the
eight amplitudes $A_{1i}$ and $A_{2i}$ and possibly other `nuisance' parameters
like
$\phi_a = \Omega t_a$. Instead, in practice to date,
frequency-domain searches for binary
sources---including in this paper---seek to sum the \fstat{} power in
orbital sidebands efficiently with suitable weightings
in order to concentrate the
signal power into as few frequency bins as possible. This represents a
quasi--maximum-likelihood approach, because the \fstat{} maximises
$\Lambda_1'$ for each sideband \emph{separately}; as noted in
Section~\ref{ssec:bstat}, this procedure implicitly
picks different $A_{1i}$ and $A_{2i}$ values at each sideband
[viz. $A_{1i}^{(s)}$ and $A_{2i}^{(s)}$, where $s$ is the order of the
sideband].
A
true maximum-likelihood estimator, in contrast, maximises $\Lambda_1'$ for
all the sidebands added together for a single, optimal set of eight amplitudes
$\{A_{1i}, A_{2i}\}$.

In this appendix, we review two quasi--maximum-likelihood
estimators, which are independent of orbital phase, namely the
\cstat{} \citep{2007messenger,2014sammut} and Bessel-weighted
\fstat{} \citep{2016suvorova}. We then examine how to refine these estimators
to include the orbital phase.

The \cstat{} weights the power in the central $M = 2\mathrm{ceil}(2\pi f_0 a_0)
+ 1$
orbital sidebands equally without any phase correction:
\begin{align}
	\mathcal{C}(f) = \sum_{s=-(M-1)/2}^{(M-1)/2} \mathcal{F}(f - s/P).
	\label{eqn:appcstat}
\end{align}
Here $\mathrm{ceil}(...)$ returns the smallest integer greater than or equal to
its argument, and $P$ and $a_0$ denote the orbital period and light travel time
across the projected semimajor axis respectively.
The Bessel-weighted \fstat{} weights the power in the central $M$ orbital
sidebands by the squared amplitude of the Bessel envelope of the Fourier
decomposition of a frequency modulated harmonic signal \citep{2016suvorova}:
\begin{align}
	G(f) = \sum_{s=-(M-1)/2}^{(M-1)/2} J_s^2(2\pi f a_0) \mathcal{F}(f - s/P).
	\label{eqn:appbesselweight}
\end{align}
Here $J_s$ denotes a Bessel function of order $s$ of the first kind.
Equation~(\ref{eqn:appbesselweight}), like~(\ref{eqn:appcstat}), does not
exploit the information contained in the Fourier phases of the orbital
sidebands; it is a sum of real-valued, positive terms. In
Ref.~\citep{2016suvorova},~(\ref{eqn:appbesselweight}) is evaluated in practice by
first convolving the SFT data with a Bessel filter [see equations~(37) and~(38) of
the latter reference] constructed for the average $f$ in a 1-Hz sub-band
(instead of separately for every individual frequency bin) to save
computational cost.

To generalise $\mathcal{C}(f)$ and $G(f)$ to include orbital phase, we
expand~(\ref{eqn:ht})--(\ref{eqn:h24}) with the phase
model~(\ref{eqn:phasemodel})
as a
Jacobi-Anger
sum of orbital sidebands, in order to construct a signal template $h(t)$.
The result is 
\begin{align}
	h(t) = \sum_{i=1}^4 \sum_{s=-\infty}^{\infty} (-1)^s
	  J_s(2\pi f_0 a_0) A_{1i} h_{1i}^{(s)}(t),
	\label{eqn:a:htjacobi}
\end{align}
with
\begin{align}
	h_{11}^{(s)}(t) &= a(t)\cos(2\pi f_0 t + s\Omega t - s\phi_a), \\
	h_{12}^{(s)}(t) &= b(t)\cos(2\pi f_0 t + s\Omega t - s\phi_a), \\
	h_{13}^{(s)}(t) &= a(t)\sin(2\pi f_0 t + s\Omega t - s\phi_a), \\
	h_{14}^{(s)}(t) &= b(t)\sin(2\pi f_0 t + s\Omega t - s\phi_a),
\end{align}
where $f_0$ is the gravitational wave search frequency, and we write $\phi_a =
\Omega t_a$. In general $h(t)$ contains components with $f_0=f_\star$
[amplitudes $A_{1i}$ in~(\ref{eqn:ht})] and $f_0 = 2f_\star$ [amplitudes
$A_{2i}$ in~(\ref{eqn:ht})]. The latter components lead to analogous terms
in~(\ref{eqn:a:htjacobi}) involving analogous factors $h_{2i}^{(s)}(t)$, with
$f_0$ replaced by $2f_0$, which can be added easily if required.

The quasiharmonic functions $h_{1i}^{(s)}(t)$ involve a rapid oscillation at
frequency $f_0 + s/P$ modulated by a slow, diurnal oscillation introduced by the
beam pattern functions $a(t)$ and $b(t)$ defined in Ref.~\citep{1998jks}. They satisfy the
following orthogonality relation with respect to the inner
product~(\ref{eqn:innerproduct}):
\begin{align}
	(h_{1i}^{(s)} || h_{1j}^{(s')} ) = \frac{1}{2}H_{ij}\delta_{s,s'},
	\label{eqn:a:orthog}
\end{align}
with
\begin{align}
	H_{ij} = \left(
	\begin{matrix}
		A & C & 0 & 0 \\
		C & B & 0 & 0 \\
		0 & 0 & A & C \\
		0 & 0 & C & B
	\end{matrix}
	\right).
	\label{eqn:a:hijmatrix}
\end{align}
Equation~(\ref{eqn:a:orthog}) holds because (i) we
truncate the sum over Bessel orders in~(\ref{eqn:a:htjacobi}) to $M$ terms as
in~(\ref{eqn:appcstat}) and~(\ref{eqn:appbesselweight}),
yielding $|s\Omega| < 2\pi f_0
a_0 \Omega \ll 2\pi f_0$ for all $s$ (e.g., $a_0\Omega = 1.33\times10^{-4}$ for
Sco~X-1), so that even widely separated Bessel
orders are orthogonal; and (ii) we have $\Omega \Tdrift \gtrsim 10$ typically
(e.g. $\Omega \Tdrift = 79.8$ with $\Tdrift = 10\,\mathrm{d}$ for Sco~X-1),
so that beats between neighbouring Bessel orders $s' = s \pm
1$, $s \pm 2$, ... are integrated over $\gtrsim 10$ cycles in the inner
product and therefore `wash out'.

We compute the log likelihood from~(\ref{eqn:dfnlambda}) and
(\ref{eqn:a:htjacobi})--(\ref{eqn:a:hijmatrix}) in the usual way. The result is
\begin{align}
	\Lambda'_1 &= (x || h) - \tfrac{1}{2}( h || h )
	 \label{eqn:a:lambdaprimesimple}
	\\
	 &= \sum_{i=1}^4 \sum_{s=-\infty}^\infty (-1)^s J_s(2\pi f_0 a_0)
	  \tilde{A}_{1i}^{(s)} (x || h_{1i}^{(s)}|_{\phi_a = 0})
	  \notag \\
	  &\phantom{=} -\frac{1}{4} \sum_{i,j = 1}^4 \sum_{s=-\infty}^{\infty}
	  [J_s(2\pi f_0 a_0)]^2 \tilde{A}^{(s)}_{1i} \tilde{A}^{(s)}_{1j} H_{ij},
	 \label{eqn:a:lambdaprime}
\end{align}
with
\begin{align}
	\tilde{A}^{(s)}_{11} &= A_{11} \cos s\phi_a - A_{13} \sin s\phi_a, \\
	\tilde{A}^{(s)}_{12} &= A_{12} \cos s\phi_a - A_{14} \sin s\phi_a, \\
	\tilde{A}^{(s)}_{13} &= A_{11} \sin s\phi_a + A_{13} \cos s\phi_a, \\
	\tilde{A}^{(s)}_{14} &= A_{12} \sin s\phi_a + A_{14} \cos s\phi_a.
	  \label{eqn:a:Atildes24}
\end{align}
In writing (\ref{eqn:a:lambdaprime})--(\ref{eqn:a:Atildes24}), we transfer the
unknown $\phi_a$ out of the inner product, leaving $(x || h_{1i}^{(s)}|_{\phi_a
= 0})$, and fold it into the coefficients $\tilde{A}_{1i}$. Thus $(x ||
h_{1i}^{(s)}|_{\phi_a = 0})$ is independent of $\phi_a$ and can be computed
from the data stream $x(t)$ using the standard \fstat{} given $f_0$ and $P$.

Suppose we now seek to maximise~(\ref{eqn:a:lambdaprime}) with respect to the five
unknowns $A_{11}$, $A_{12}$, $A_{13}$, $A_{14}$, and $\phi_a$.
This leads to five nonlinear, simultaneous equations, each containing $M$ terms
from the truncated Bessel sums.
The equations are poorly conditioned, because the terms oscillate rapidly as
functions of $\phi_a$ with periods $2\pi, \pi, ..., 4\pi / (M-1)$.
It is therefore tempting to maximise each Bessel
order separately by way of approximation,
as we do implicitly in~(\ref{eqn:appcstat})
and~(\ref{eqn:appbesselweight}). Writing $\Lambda'_1 = \sum_s
\Lambda_1^{\prime(s)}$, we observe that $\Lambda_1^{\prime(s)}$ is linear in
$A_{1i}$. The linear subsystem $(1 \leq i \leq 4)$
\begin{align}
	0 = \frac{\partial\Lambda_{1}^{\prime(s)}}
	   {\partial A_{1i}^{(s)}}
	   \label{eqn:a:maximise}
\end{align}
can be solved to give
\begin{align}
	A_{11}^{(s)} &= 2D^{-1} \{ [ B (x || h_{11}^{(s)}|_{\phi_a = 0} )
	 - C (x || h_{12}^{(s)}|_{\phi_a = 0}) ] \cos \theta \notag \\
	  & \phantom{=} + [B (x || h^{(s)}_{13}|_{\phi_a = 0} )
	 - C (x || h_{14}^{(s)}|_{\phi_a = 0}) ] \sin\theta \} \label{eqn:a:a21s} \\
	A_{12}^{(s)} &= 2D^{-1} \{ [ A (x || h_{12}^{(s)}|_{\phi_a = 0} )
	 - C (x || h_{11}^{(s)}|_{\phi_a = 0}) ] \cos \theta \notag \\
	  & \phantom{=} + [A (x || h^{(s)}_{14}|_{\phi_a = 0} )
	 - C (x || h_{13}^{(s)}|_{\phi_a = 0}) ] \sin\theta \} \\
	A_{13}^{(s)} &= 2D^{-1} \{ [ B (x || h_{13}^{(s)}|_{\phi_a = 0} )
	 - C (x || h_{14}^{(s)}|_{\phi_a = 0}) ] \cos \theta \notag \\
	  & \phantom{=} + [C (x || h^{(s)}_{12}|_{\phi_a = 0} )
	 - B (x || h_{11}^{(s)}|_{\phi_a = 0}) ] \sin\theta \} \\
	A_{14}^{(s)} &= 2D^{-1} \{ [ A (x || h_{14}^{(s)}|_{\phi_a = 0} )
	 - C (x || h_{13}^{(s)}|_{\phi_a = 0}) ] \cos \theta \notag \\
	  & \phantom{=} + [C (x || h^{(s)}_{11}|_{\phi_a = 0} )
	 - A (x || h_{12}^{(s)}|_{\phi_a = 0}) ] \sin\theta \}, \label{eqn:a:a24s}
\end{align}
for each $s$, with $\theta = s\phi_a$,
where $A_{1i}^{(s)}$ denotes the amplitude $A_{1i}$ in
$\Lambda_1^{\prime(s)}$ (see first paragraph of this appendix).
Upon
substituting~(\ref{eqn:a:a21s})--(\ref{eqn:a:a24s}) into
$\Lambda_1^{\prime(s)}$, we find that $\Lambda_1^{\prime(s)}$ is independent of
$\theta$, i.e. $\Lambda_1'$
has maximum value
\begin{align}
	\Lambda_1' &= \sum_{s=-\infty}^\infty [2 (-1)^s J_s(2\pi f_0 a_0)
	  -J_s^2(2\pi f_0 a_0) ] D^{-1} \notag \\
	 &\phantom{=} \times [B(x || h_{11}^{(s)}|_{\phi_a = 0} )^2
	  - 2C (x || h_{11}^{(s)}|_{\phi_a = 0} )(x || h_{12}^{(s)}|_{\phi_a = 0})
	   \notag \\
	 &\phantom{=} + A (x || h_{12}^{(s)}|_{\phi_a = 0} )^2 \notag \\
	 &\phantom{=} + B(x || h_{13}^{(s)}|_{\phi_a = 0} )^2
	  - 2C (x || h_{13}^{(s)}|_{\phi_a = 0} )(x || h_{14}^{(s)}|_{\phi_a = 0})
	   \notag \\
	 &\phantom{=} + A (x || h_{14}^{(s)}|_{\phi_a = 0} )^2 ] \notag \\
	   \label{eqn:a:lambda1full}
	   \\
	&= \sum_{s = -\infty}^{\infty} [2(-1)^s J_s(2\pi f_0 a_0)
	 - J_s^2(2\pi f_0 a_0) ] \mathcal{F}(f_0 + s/P).
	   \label{eqn:a:lambda1simpl}
\end{align}
The sideband phases enter~(\ref{eqn:a:lambdaprime})
through $\tilde{A}_{1i}$. They are missing from~(\ref{eqn:a:lambda1simpl})
following the approximate maximisation step in~(\ref{eqn:a:maximise}).
Hence~(\ref{eqn:a:lambda1simpl}) does not exploit the information in the
orbital phase; it does not combine the sidebands coherently.

An
alternative approach involves replacing $A_{1i}$ [\emph{not}
$\tilde{A}_{1i}^{(s)}$] by the maximum-likelihood expressions from the classic
\fstat{} in each sideband separately, i.e. replace $A_{1i}$
in~(\ref{eqn:a:htjacobi}) with $A^{(s)}_{1i}$ as given by
\begin{align}
	A_{11}^{(s)} &= 2D^{-1}[ B (x || h_{11}^{(s)}|_{\phi_a = 0} ) - C (x ||
	  h_{12}^{(s)}|_{\phi_a = 0}) ] \\
	A_{12}^{(s)} &= 2D^{-1}[ A (x || h_{12}^{(s)}|_{\phi_a = 0} ) - C (x ||
	  h_{11}^{(s)}|_{\phi_a = 0}) ] \\
	A_{13}^{(s)} &= 2D^{-1}[ B (x || h_{13}^{(s)}|_{\phi_a = 0} ) - C (x ||
	  h_{14}^{(s)}|_{\phi_a = 0}) ] \\
	A_{14}^{(s)} &= 2D^{-1}[ A (x || h_{14}^{(s)}|_{\phi_a = 0} ) - C (x ||
	  h_{13}^{(s)}|_{\phi_a = 0}) ].
\end{align}
One then evaluates~(\ref{eqn:a:lambdaprimesimple}) on a grid of $\phi_a$ values
and picks out the maximum ``by brute force'', without attempting to maximise
$\Lambda_1'$ analytically with respect to $\phi_a$. This approach leads to the
\bstat{} introduced in section~\ref{ssec:bstat}. Expressed in terms of the
Fourier integrals $\mathcal{F}_{1a}$ and $\mathcal{F}_{1b}$ which enter the
\fstat{}, it takes the form
\begin{align}
	\mathcal{J} = \frac{4}{S_h(f_0)\Tobs D} \left[ B|\mathcal{J}_{1a}|^2
	- 2C \mathrm{Re}(\mathcal{J}_{1a}\mathcal{J}_{1b}^\star)
	+ A|\mathcal{J}_{1b}|^2 \right],
	\label{eqn:a:bstat}
\end{align}
where $\mathcal{J}_{1a}$ and $\mathcal{J}_{1b}$ are given by
\begin{align}
	\mathcal{J}_{1a} &=
	\sum_{s = -\infty}^{\infty} J_s(2\pi f_0 a_0) e^{-is\phi_a} \mathcal{F}_{1a}(f_0 + s/P), \\
	\mathcal{J}_{1b} &=
	\sum_{s = -\infty}^{\infty} J_s(2\pi f_0 a_0) e^{-is\phi_a} \mathcal{F}_{1b}(f_0 + s/P).
\end{align}
Note that~(\ref{eqn:a:bstat}) is still an approximate,
quasi-maximum--likelihood formula for the reasons discussed in the first
paragraph of this appendix; it takes a maximum likelihood approach to every
sideband separately rather than maximising the sum over sidebands
\textit{in toto}.

The relative performances of the Bessel-weighted \fstat{} and the
orbital-phase-independent statistic~(\ref{eqn:a:lambda1simpl}) are displayed in
Figure~\ref{fig:a:fppstat}. The right panel plots $\Lambda_1'$
in~(\ref{eqn:a:lambda1simpl}) versus observing frequency $f_0$
for an injected signal with $h_0 =
8 \times 10^{-26}$ and other parameters copied from
Table~\ref{tbl:synthparams}.
$\Lambda_1'$ displays a
double-horned structure that is
similar to (albeit narrower than)
the \fstat{}.
The double-horn in the original \fstat{} has a width of $2.91 \times
10^{-2}\,\mathrm{Hz}$; the estimator~(\ref{eqn:a:lambda1simpl})
narrows this to $6.22
\times 10^{-3}\,\mathrm{Hz}$, although it remains wider than the full-width
half-maximum of the Bessel-weighted \fstat{}~(\ref{eqn:appbesselweight})
($2.93\times10^{-4}\,\mathrm{Hz}$; see Figure~\ref{fig:comparison}).
Both~(\ref{eqn:a:lambda1simpl}) and
the
Bessel-weighted \fstat{}, peak $\approx 14\,\mathrm{dB}$ above the noise.
The \bstat{} does even
better. In the bottom right panel of Figure~\ref{fig:comparison}, the signal is
concentrated entirely into one frequency bin at $f = f_\star$; there are
no shoulders around the peak, unlike~(\ref{eqn:appbesselweight})
and~(\ref{eqn:a:lambda1simpl}),
and the \bstat{} peaks $32\,\mathrm{dB}$ above the noise.

\begin{figure}
	\includegraphics[width=0.49\textwidth]{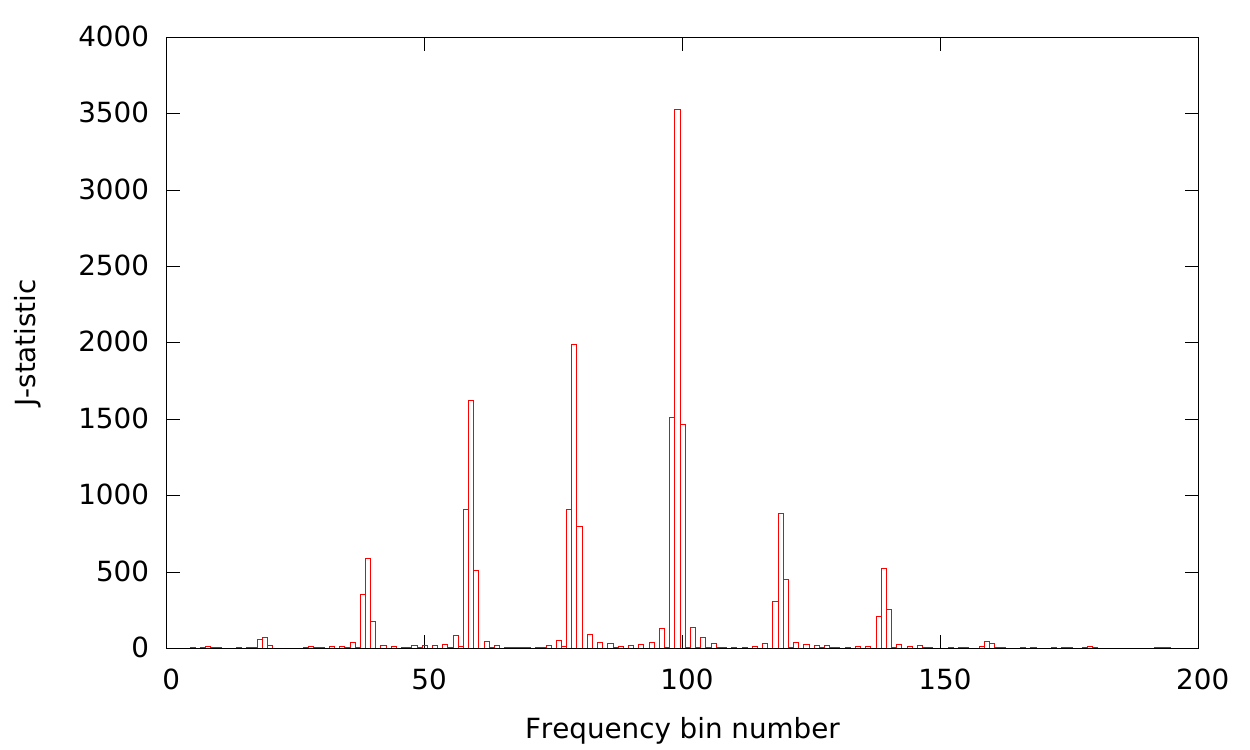}
\caption{
	Zoomed-in portion of the \bstat{} for an injection with
	$h_0 = 8\times10^{-25}$, showing diurnal
	sidebands (six clearly discernible).
	Each frequency bin covers
	$\Delta \fdrift = 5.787\times 10^{-7}\,\mathrm{Hz}$. Frequency bin number
	100 corresponds to the injection frequency $f_\star = 111.1\,\mathrm{Hz}$.
	The sidebands are
	$\approx 20$ frequency bins apart, i.e., $1/(86400\,\mathrm{s})$.
	}
\label{fig:a:bstatdiurnal}
\end{figure}

The \bstat{}, like the \fstat{}, leaks a small amount of residual
power into nonorbital sidebands bracketing the central spectral line.
Figure~\ref{fig:a:bstatdiurnal}
displays a close-up of the \bstat{} for a strong injection, with $h_0 =
8\times10^{-25}$. Individual frequency bins are discernible across
a $1.1574\times10^{-4}\,\mathrm{Hz}$ band.
The
sharp peak coincident with the injected signal splits into sidebands spaced by
approximately 20 (out of 200) frequency bins, i.e. $1.157\times10^{-5}\,\mathrm{Hz} =
1/(86400\,\mathrm{s})$. The sidebands are associated with the component of
the Earth's diurnal motion that
is not completely removed in the \fstat{}, due to approximations like the one
leading to~(\ref{eqn:f2aapprox}) and~(\ref{eqn:ahat}).
Each diurnal sideband spreads across several adjacent bins; its profile is a
$\sinc$ function produced by the observing window $\Tobs$.
The signature in Figure~\ref{fig:a:bstatdiurnal} is also
observed in the \fstat{} for an isolated source.

\section{Cram\'er-Rao lower bound for the \bstat{}}
\label{app:crlb}
In general, the Cram\'er-Rao lower bound (CRLB) of a model parameter $\theta$
estimated from
noisy measurements is the minimum possible variance of any unbiased estimator
of $\theta$. The CRLB depends on the PDF of the observed data,
specifically its curvature in the neighbourhood of the true value of $\theta$.
The more
sensitively the PDF depends on the parameter
(the greater the curvature, in other words),
the more accurately the parameter
can be estimated.

Let $x(t) = h(t; \bm{\theta}) + n(t)$ represent the output from
a single
interferometer in a gravitational-wave observatory, where $h(t; \bm{\theta})$ is the signal defined by
(\ref{eqn:ht})--(\ref{eqn:h24}), $n(t)$ represents stationary Gaussian noise,
and $\bm{\theta} = (f_0, a_0, \phi_a = \Omega t_a)$ is a vector containing
the three unknown signal parameters in the search. Let
$p(\mathbf{x};\bm{\theta})$ be the PDF of the observed data; here,
$\mathbf{x}$ is a vector containing every sample $x(0)$, $x(t_1)$, $x(t_2)$,
..., $x(\Tobs)$ of the interferometer output over the full observation $0 \leq
t_i \leq \Tobs$. We define the Fisher information matrix $\mathbf{I}$ by its
entries
\begin{align}
	I_{ij} = -\left\langle \frac{\partial^2 \ln p}{\partial \theta_i \partial
	\theta_j} \right\rangle,
	\label{eqn:b:fisherIij}
\end{align}
where $\langle \cdots \rangle$ denotes the expectation value taken over many
realisations of the noise. Then the CRLB for the parameter $\theta_i$ is
\citep{1993kay}
\begin{align}
	\mathrm{var}(\theta_i) \geq (\mathbf{I}^{-1})_{ii},
	\label{eqn:b:varcond}
\end{align}
where $\mathbf{I}^{-1}$ is the matrix inverse of $\mathbf{I}$, and
$(\mathbf{I}^{-1})_{ii}$ symbolises the $i$-th diagonal entry of
$\mathbf{I}^{-1}$ as opposed to its trace, i.e. the Einstein summation
convention does \textit{not} apply in~(\ref{eqn:b:varcond}).

For stationary, Gaussian noise with normalised unit variance, the log
likelihood is given by (see Section~\ref{sec:matchedfilter})
\begin{align}
	\ln p = -\frac{1}{2}(x - h || x - h)
	\label{eqn:b:logep}
\end{align}
up to a constant.
The model $h(t; \bm{\theta})$ depends on $\bm{\theta}$, and the
inner product~(\ref{eqn:innerproduct}) in~(\ref{eqn:b:logep}) is symmetric, so
the derivatives of the log likelihood reduce to
\begin{align}
	\frac{\partial \ln p}{\partial \theta_i} =
	\left(x \dblbar \frac{\partial h}{\partial \theta_i} \right)
	 - \left(h \dblbar \frac{\partial h}{\partial \theta_i} \right)
\end{align}
and
\begin{align}
	\frac{\partial^2 \ln p}{\partial\theta_i \partial\theta_j} =
	-\left(\frac{\partial h}{\partial \theta_i} \dblbar \frac{\partial
	h}{\partial\theta_j} \right)
	+ \left(x - h \dblbar \frac{\partial^2h}{\partial\theta_i \partial\theta_j}
	\right).
	\label{eqn:b:fisher2d}
\end{align}
When the ensemble average is taken, the second term in~(\ref{eqn:b:fisher2d})
vanishes, because one has $\langle x - h \rangle = 0$ for an unbiased
estimator. Equations~(\ref{eqn:b:fisherIij}) and~(\ref{eqn:b:fisher2d}) then
imply
\begin{align}
	I_{ij} = \left( \frac{\partial h}{\partial\theta_i} \dblbar
	 \frac{\partial h}{\partial\theta_j} \right).
	 \label{eqn:b:fishercomps}
\end{align}

The derivatives $\partial h / \partial \theta_i$ are straightforward to
evaluate in terms of the signal defined by
(\ref{eqn:ht})--(\ref{eqn:h24}).
Defining $\Phi_\pm(t) = \Phi(t) \pm (\Omega t - \phi_a)$, we obtain
\begin{align}
	\frac{\partial h}{\partial \theta_i} = \sum^4_{j=1} A_{1j} \frac{\partial
	h_{1j}}{\partial \theta_i}
	\label{eqn:b:zerofisher}
\end{align}
with
\begin{align}
	\frac{\partial h_{11}}{\partial f_0} &=
	 -2\pi t a(t) \sin\Phi(t) \notag \\
	  &\phantom{=\,}- \pi a_0 a(t) [ \cos\Phi_+(t) - \cos\Phi_-(t) ],
	 \label{eqn:b:firstfisher}
	\\
	\frac{\partial h_{13}}{\partial f_0} &=
	 2\pi ta(t)\cos\Phi(t) \notag \\
	 &\phantom{=\,}- \pi a_0 a(t)[\sin\Phi_+(t) - \sin\Phi_-(t) ],
	 \label{eqn:b:fisher2}
	\\
	\frac{\partial h_{11}}{\partial a_0} &=
	 -\pi f_0 a(t) [\cos\Phi_+(t) - \cos\Phi_-(t)],
	\\
	\frac{\partial h_{13}}{\partial a_0} &=
	 -\pi f_0 a(t) [\sin\Phi_+(t) - \sin\Phi_-(t)],
	\\
	\frac{\partial h_{11}}{\partial \phi_a} &=
	 -\pi f_0 a_0 a(t) [\sin\Phi_+(t) + \sin\Phi_-(t)],
	\\
	\frac{\partial h_{13}}{\partial \phi_a} &=
	 \pi f_0 a_0 a(t) [\cos\Phi_+(t) + \cos\Phi_-(t)].
	 \label{eqn:b:lastfisher}
\end{align}
In the same way $\partial h_{12}/\partial f_0$, $\partial h_{14}/\partial f_0$,
$\partial h_{12} / \partial a_0$, $\partial h_{14} / \partial a_0$, $\partial
h_{12} / \partial \phi_a$, and $\partial h_{14} / \partial \phi_a$ are obtained
by replacing $a(t)$ with $b(t)$
in~(\ref{eqn:b:firstfisher})--(\ref{eqn:b:lastfisher}) respectively.

When evaluating $I_{ij}$ from~(\ref{eqn:b:fishercomps})
using~(\ref{eqn:b:zerofisher})--(\ref{eqn:b:lastfisher}), we note three points.
(i) In $\partial h_{1j}/\partial f_0$, the first terms on the right-hand sides
of~(\ref{eqn:b:firstfisher}) and~(\ref{eqn:b:fisher2}) are larger
than the second and third terms
by a factor
of $\approx 2t / a_0 \gg 1$, implying
$\partial h_{11} / \partial f_0 \approx -2\pi t a(t)\sin\Phi(t)$ and $\partial
h_{13} / \partial f_0 \approx 2\pi t a(t) \cos\Phi(t)$.
For example, we have $a_0 = 1.44\,\mathrm{s}$ and $t \leq \Tdrift = 10\,\mathrm{d}$
for Sco X-1.
(ii) The beam-pattern functions $a(t)$ and $b(t)$ oscillate about nonzero
means. Specifically they are linear combinations of DC terms and sinusoids with
periods of $0.5\,\mathrm{d}$ and $1.0\,\mathrm{d}$; see equations~(12) and~(13)
in Ref.~\citep{1998jks}. As the latter periods are typically much shorter than
$\Tdrift$, the relevant timespan for calculating the \bstat{}, we can
write $(ta || ta) = \frac{1}{3}\Tdrift^2 A$, $(tb||tb) = \frac{1}{3} \Tdrift^2 B$
and $(ta || tb) = \frac{1}{3}\Tdrift^2 C$ plus correction terms of order
$(\Tdrift / 1\,\mathrm{d})^{-1}$, with $A$, $B$, and $C$ defined
following equation~(\ref{eqn:innerproduct}).
(iii) The off-diagonal products $(\partial h / \partial \theta_i)(\partial h /
\partial \theta_j)$ with $i \neq j$ are composed of linear combinations of
terms oscillating harmonically in time with zero means. Again the
oscillation periods are typically much shorter than $\Tdrift$,
yielding
$I_{ij} = 0$ for $i \neq j$ to a good
approximation [plus correction terms of order $\mathrm{max}(P, 1\,\mathrm{d}) /
\Tdrift \ll 1$]. For example, $(\partial h / \partial f_0) (\partial h / \partial
a_0)$ is a linear combination of terms proportional to $\sin(\Omega t -
\phi_a) \cos^2 \Phi(t)$, $\sin(\Omega t - \phi_a)\cos\Phi(t)\sin\Phi(t)$ and
$\sin(\Omega t - \phi_a) \sin^2\Phi(t)$, which oscillate proportional
to $\exp[\pm 2 i \Phi(t) \pm i \Omega t]$ and $\exp(\pm i\Omega t)$ when
expanded. Likewise,
$(\partial h / \partial a_0) (\partial h/\partial \phi_a)$ is a linear
combination of terms proportional to $\sin 2(\Omega t - \phi_a)$ multiplied by
$\cos^2\Phi(t)$, $\cos\Phi(t)\sin\Phi(t)$, and $\sin^2\Phi(t)$,
which
oscillate
proportional to $\exp[\pm2i\Phi(t) \pm 2i\Omega t]$ and $\exp(\pm
2i\Omega t)$ when expanded.

Putting together points (i)--(iii) above, we find that the Fisher information
matrix is approximately diagonal, i.e. $I_{ij} \approx
\mathrm{diag}(I_{f_0f_0}, I_{a_0a_0}, I_{\phi_a\phi_a})$, with
\begin{align}
	I_{f_0f_0} &= \frac{2 \Tobs^2 I_{a_0a_0}}{3 f_0^2}, \\
	I_{a_0a_0} &= \pi^2 f_0^2 [A(A_{11}^2 + A_{13}^2)
	 + B(A_{12}^2 + A_{14}^2) \notag \\
	 &\phantom{=}
	 + 2C(A_{11}A_{12} + A_{13}A_{14}) ],
	 \label{eqn:b:ia0a0}
	 \\
	I_{\phi_a\phi_a} &= a_0^2 I_{a_0a_0}.
\end{align}
The factor $[\cdots]$ in square brackets in~(\ref{eqn:b:ia0a0}) equals twice the
noncentrality parameter $\lambda$ appearing in the chi-squared
PDF of the \fstat{}, i.e., $p(2\mathcal{F}) = \chi^2(2\mathcal{F}; 4,
\lambda)$; see Section~III$\,$A in Ref.~\citep{2016suvorova}. The CRLBs on the three
parameters follow directly from~(\ref{eqn:b:varcond}):
\begin{align}
	\mathrm{var}(f_0) &\geq \frac{3f_0^2}{2\Tobs^2 I_{a_0a_0}}, \\
	\mathrm{var}(a_0) &\geq \frac{1}{I_{a_0a_0}}, \\
	\mathrm{var}(\phi_a) &\geq \frac{1}{a^2_0 I_{a_0a_0}}.
\end{align}

\section{False alarm and dismissal rates}
\label{app:far}

\begin{figure}
	\includegraphics[width=0.49\textwidth]{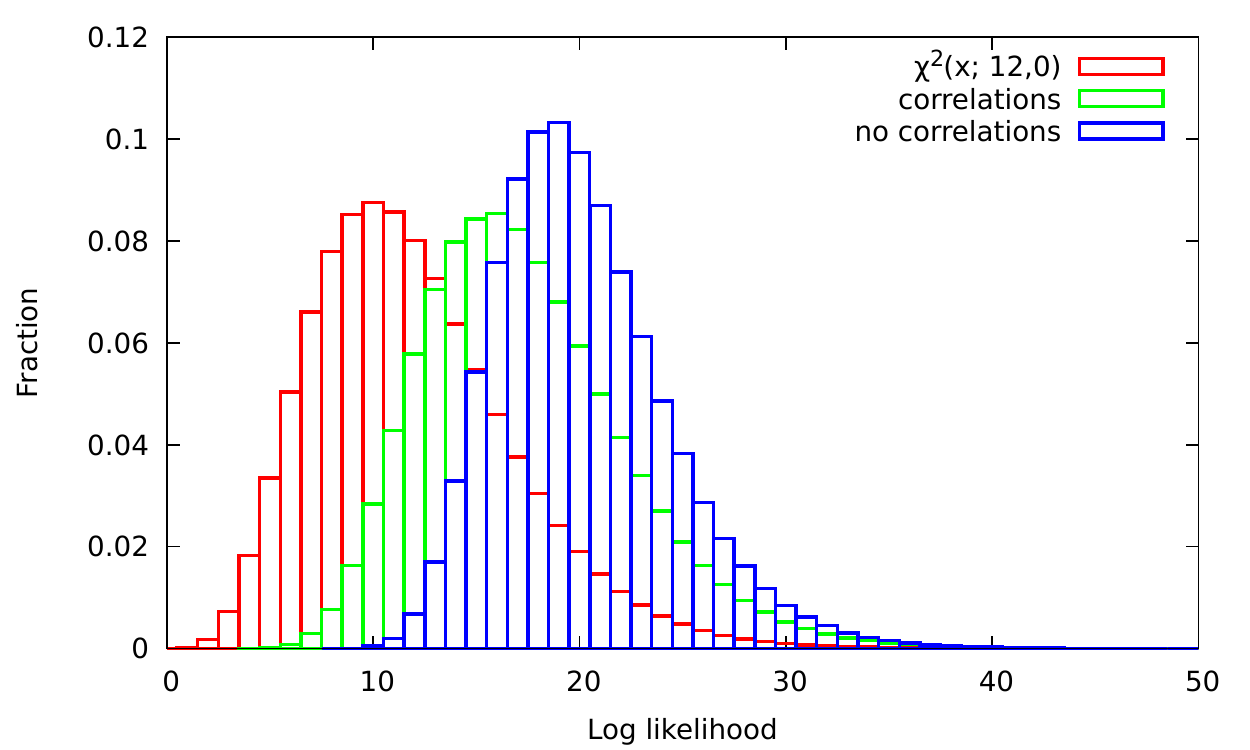}
	\caption{Correlations and maximisation in the HMM log likelihood statistics
	for
	the illustrative, three-step example in Appendix~\ref{app:far}:
	PDF of the sum of three $\mathcal{F}$- or \bstat{} values (red
	histogram); maximum log likelihood of nine arbitrary,
	independent Viterbi paths (blue curve);
	maximum log likelihood of the nine Viterbi paths in a single realisation of
	synthetic noise (green curve).}
	\label{fig:pdfhistog}
\end{figure}

\begin{figure}
	\includegraphics[width=0.49\textwidth]{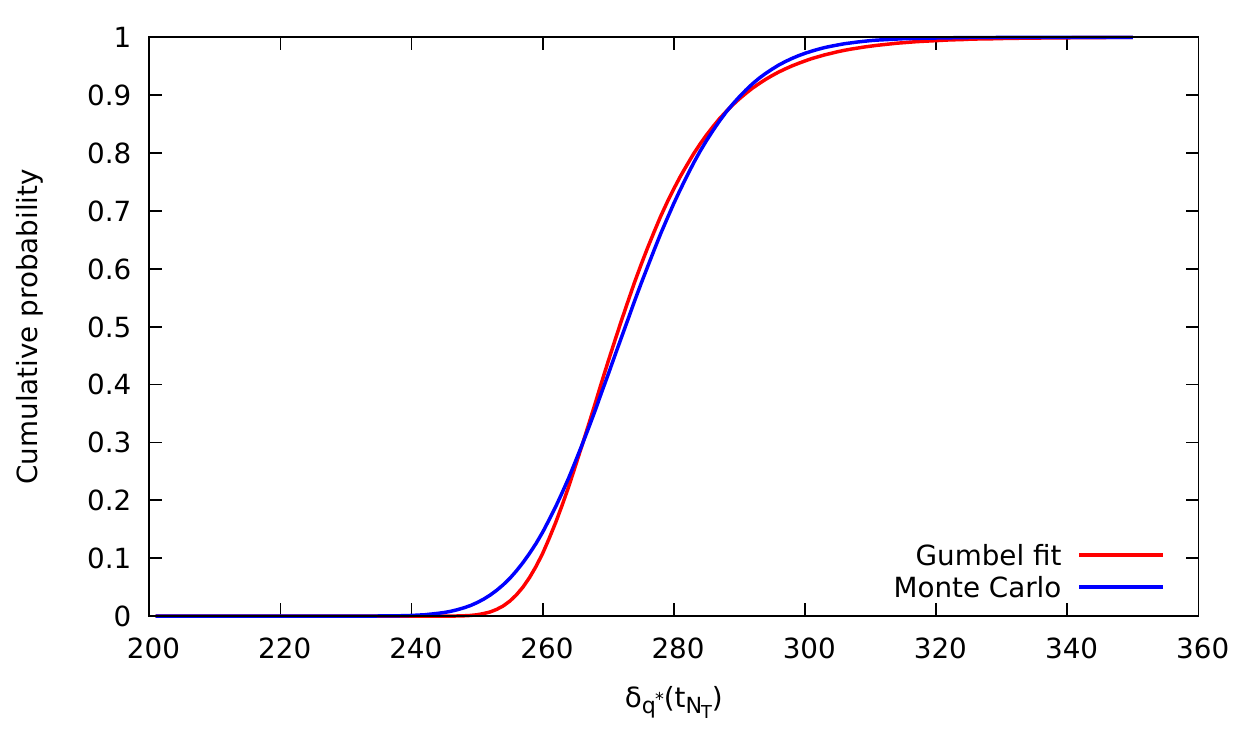}
	\caption{Least-squares fit of a Gumbel law (\ref{eqn:kest}) (red curve)
	to $K(z, \lambda = 0)$ in (\ref{eqn:c:fz}) derived
	from Monte-Carlo simulations (blue curve)
	for the representative
	example $N_T = 37$ and $N_Q =
	1.73\times10^6$. The best fit is for $a = 10.0$ and $b = 268$.
	}
	\label{fig:rmse:a}
\end{figure}

\begin{table*}
	\caption{Gumbel law [see~(\ref{eqn:kest})] parameters $a(N_T, N_Q)$ and
	$b(N_T, N_Q)$ for empirical fits to $K(z, \lambda = 0)$ as functions of
	$N_T$ and $N_Q$ in ranges useful in practice. The RMSE column gives the
	root-mean-square error between the fit and empirical cumulative
	distribution function from Monte-Carlo simulations; one finds RMSE $< 1\%$.
	The values of $N_Q$ are equally spaced logarithmically.}
	\label{tab:c:rmse}
	\begin{tabular}{r@{\hspace{5mm}}rrr@{\hspace{5mm}}rrr@{\hspace{5mm}}rrr} \hline
		& \multicolumn{3}{c}{$N_Q = 1.73\times 10^6$} & \multicolumn{3}{c}{$N_Q = 1.73\times 10^6/\sqrt{10}$} & \multicolumn{3}{c}{$N_Q = 1.73\times 10^6 / 10$} \\
		$N_T$ & $a$ & $b$ & RMSE & $a$ & $b$ & RMSE & $a$ & $b$ & RMSE \\
		\hline
 1 & $2.13$ & $32.13$  & $7.8 \times 10^{-3}$ & $2.13$ & $32.13$ &  $7.8 \times 10^{-3}$ & $2.13$ & $32.13$ &  $7.8 \times 10^{-3}$ \\
 5 & $2.67$ & $73.83$  & $6.5 \times 10^{-3}$ & $2.67$ & $73.83$ &  $6.5 \times 10^{-3}$ & $2.67$ & $73.83$ &  $6.5 \times 10^{-3}$ \\
10 & $3.35$ & $119.27$ & $6.0 \times 10^{-3}$ & $3.35$ & $119.27$ & $6.0 \times 10^{-3}$ & $3.35$ & $119.27$ & $6.0 \times 10^{-3}$ \\
15 & $3.46$ & $162.99$ & $5.7 \times 10^{-3}$ & $3.46$ & $162.99$ & $5.7 \times 10^{-3}$ & $3.46$ & $162.99$ & $5.7 \times 10^{-3}$ \\
20 & $3.44$ & $205.59$ & $5.5 \times 10^{-3}$ & $3.44$ & $205.59$ & $5.5 \times 10^{-3}$ & $3.44$ & $205.59$ & $5.5 \times 10^{-3}$ \\
25 & $3.75$ & $247.62$ & $5.3 \times 10^{-3}$ & $3.75$ & $247.62$ & $5.3 \times 10^{-3}$ & $3.75$ & $247.62$ & $5.3 \times 10^{-3}$ \\
30 & $4.25$ & $289.34$ & $5.1 \times 10^{-3}$ & $4.25$ & $289.34$ & $5.1 \times 10^{-3}$ & $4.25$ & $289.34$ & $5.1 \times 10^{-3}$ \\
37 & $4.55$ & $347.18$ & $4.6 \times 10^{-3}$ & $4.55$ & $347.18$ & $4.6 \times 10^{-3}$ & $4.55$ & $347.18$ & $4.6 \times 10^{-3}$ \\
\hline
	\end{tabular}
\end{table*}

In order to calculate the false alarm probability $\falsealarm$ and false
dismissal probability
$\falsedismissal$ for the algorithm developed in this paper, one needs the PDFs of the
Viterbi probabilities after $k$ steps of the HMM in the absence and presence of
a
signal respectively. As far as the authors know, no general formula for these
HMM
PDFs exists in the literature for a chi-squared--distributed estimator like the
\fstat{} or \bstat{}. In this appendix, we review why the HMM maximisation step
makes it hard to calculate the score PDF (section~\ref{ssec:c:correlation}), present an
approximate, empirical distribution whose form is suggested by extreme value
theory
(section~\ref{ssec:c:scorepdf}), and quantify $\falsealarm$ and
$\falsedismissal$ in terms of
the empirical distribution
(section~\ref{ssec:c:farformula}).
We improve on a first attempt at these calculations in
Ref.~\citep{2016suvorova} and
compute the probability of outliers more realistically.

\subsection{Viterbi path correlation and maximisation}
\label{ssec:c:correlation}
Equation~(33) in Ref.~\citep{2016suvorova}
estimates $\falsealarm$ crudely
by assuming that $\max \log \Pr(Q|O)$ follows
a central chi-squared distribution with $4k$ degrees of freedom
after $k$ HMM steps, because $2\mathcal{F} = \log L_{o(t_k)q(t_k)}$
is drawn from
the PDF $p(2\mathcal{F}) = \chi^2(2\mathcal{F}; 4, 0)$
in the absence of a signal, and the chi-squared distribution is
additive. However, this assumption breaks down on two counts. First, the
nonlinear maximisation operator in the Viterbi algorithm returns values from the
\textit{tail} of $\chi^2(2\mathcal{F};4,0)$, because
$\chi^2(2\mathcal{F};4,0)$ is sampled $N_Q$ times,
once for each possible transition from the previous step.
Second, the Viterbi paths overlap
partially, so the random numbers $\log L_{o(t_j)q(t_j)}$
for $1 \leq j \leq k$ are
not independent and identically distributed.
Exactly the same issues arise, if the frequency domain estimator at each HMM
step is the \bstat{} instead of the \fstat{}.

Consider all admissible paths following the transition rule
in equation~(\ref{eqn:transmat}), that end in state $q_i$ after the $k$-th HMM
step.
Label
the log likelihood of
each path by $x_p = \log \Pr(Q|O)$, with $1 \leq p \leq 3^k$.
We wish to compute the cumulative
probability that $\max_p x_p$ is less than $z$, viz.
\begin{align}
	K(z; \lambda) = \Pr(x_1 < z, \cdots, x_{3^k} < z)
	\label{eqn:c:fz}
\end{align}
where $\lambda$ is the non-centrality parameter (zero for the case of noise,
and positive for signal plus noise) which is related to the gravitational wave
signal strength by equation~(\ref{eqn:lambdadef}) in Section~\ref{ssec:relperf}.

A difficulty arises because $x_1, ..., x_{3^k}$ are correlated,
so the joint distribution cannot be written as a product of individual
probabilities. We illustrate with an example. Consider all admissible
paths up to $k = 3$ ending in $q_3$. We have $x_1 = X(1, 1) + X(2, 2) + X(3, 3)$ for the path
$\{q_1, q_2, q_3\}$, $x_2 = X(2, 1) + X(2, 2) + X(3, 3)$ for the path $\{q_2,
q_2, q_3\}$, and so on, where $X(i, j)$ are independent samples of the \fstat{}
or \bstat{}
in state $q_i$ at the $j$-th HMM step. We can write the sums in matrix notation
as $\vect{x} = A\vect{u}$ with $\vect{x} = (x_1, \cdots, x_9)^T$, $\vect{u}
= [ X(1,1),\allowbreak X(2, 1),\allowbreak X(3, 1),\allowbreak X(4,
1),\allowbreak X(5, 1),\allowbreak X(2, 2),\allowbreak X(3, 2),\allowbreak X(4,
2),\allowbreak X(3, 3) ]^T$ and
\begin{align}
	A = \left(\begin{matrix}
		1 & 0 & 0 & 0 & 0 & 1 & 0 & 0 & 1 \\
		0 & 1 & 0 & 0 & 0 & 1 & 0 & 0 & 1 \\
		0 & 0 & 1 & 0 & 0 & 1 & 0 & 0 & 1 \\
		0 & 1 & 0 & 0 & 0 & 0 & 1 & 0 & 1 \\
		0 & 0 & 1 & 0 & 0 & 0 & 1 & 0 & 1 \\
		0 & 0 & 0 & 1 & 0 & 0 & 1 & 0 & 1 \\
		0 & 0 & 1 & 0 & 0 & 0 & 0 & 1 & 1 \\
		0 & 0 & 0 & 1 & 0 & 0 & 0 & 1 & 1 \\
		0 & 0 & 0 & 0 & 1 & 0 & 0 & 1 & 1
	\end{matrix}\right).
\end{align}

The covariance of $\vect{x}$ is
\begin{align}
	\langle x_i, x_j \rangle = AA^T = \left(\begin{matrix}
		3 & 2 & 2 & 1 & 1 & 1 & 1 & 1 & 1 \\
		2 & 3 & 2 & 2 & 1 & 1 & 1 & 1 & 1 \\
		2 & 2 & 3 & 1 & 2 & 1 & 2 & 1 & 1 \\
		1 & 2 & 1 & 3 & 2 & 2 & 1 & 1 & 1 \\
		1 & 1 & 2 & 2 & 3 & 2 & 2 & 1 & 1 \\
		1 & 1 & 1 & 2 & 2 & 3 & 1 & 2 & 1 \\
		1 & 1 & 2 & 1 & 2 & 1 & 3 & 2 & 2 \\
		1 & 1 & 1 & 1 & 1 & 2 & 2 & 3 & 2 \\
		1 & 1 & 1 & 1 & 1 & 1 & 2 & 2 & 3 \\
	\end{matrix}\right).
	\label{eqn:c:covarmat}
\end{align}
Equation~(\ref{eqn:c:covarmat}) is clearly not diagonal. At the time of
writing, it is unclear how to
fold equation~(\ref{eqn:c:covarmat}) analytically into
the computation of $K(z, \lambda = 0)$.

\subsection{Log likelihood PDF}
\label{ssec:c:scorepdf}
Although it is challenging to calculate the PDF of $\max_p x_p = \max_Q \log
\Pr(Q|O)$ theoretically, it is relatively simple, albeit time-consuming, to
compute it empirically. Figure~\ref{fig:pdfhistog}
plots
three histograms for the
illustrative example of a
three-step HMM:
the PDF of the sum of three independent $\mathcal{F}$- or \bstat{}
values, which
matches a central chi-squared distribution with 12 degrees of freedom (red
histogram); the PDF of $\delta_{q_i}(t_{N_T})$ without taking correlations into account,
i.e.
the maximum log likelihood for any nine independent
paths, each path comprising three independent $\mathcal{F}$- or \bstat{}
samples (blue histogram); and the PDF of
$\delta_{q_i}(t_{N_T})$
taking
correlations into account, i.e. the maximum log likelihood from
the nine paths in the vector $\vect{x}$ for a single realisation of a synthetic
observation (green histogram).
The PDF taking correlations into account peaks to the right of the PDF that
neglects correlations, because extreme \fstat{} values are likely to end up in
multiple paths (if they are large) or end up in few paths (if they are small).
Correlations therefore play a
significant role.

Extreme value theory states that there exist three PDF families that
describe asymptotically the maximum of $N'$ samples of a random variable for
$N' \gg 1$: the Weibull, Gumbel
and Fr\'echet laws \citep{2004sornette}. The families correspond to light,
exponential, and heavy tails respectively in the PDF of the underlying random
variable. Here we seek empirically the best fit to $K(z, \lambda)$ in~(\ref{eqn:c:fz}).
The underlying variable $x_p$ is crudely chi-squared distributed, even
when the correlations discussed in Section~\ref{ssec:c:correlation} are
included; the tail is exponential, which is easy to verify by replotting
Figure~\ref{fig:pdfhistog} on log-linear axes.
Testing by
trial and
error confirms that the Gumbel Law is a superior fit compared to the Weibull and
Fr\'echet laws, with
\begin{align}
	K(z, \lambda = 0) = \exp\{-\exp[-(z-b)/a]\},
	\label{eqn:kest}
\end{align}
where $a(N_T, N_Q)$ and $b(N_T, N_Q)$ are dimensionless parameters. Note that
\ref{eqn:kest} strictly
applies to a variable
that takes values along the whole real line.
In this application,
in contrast, we have $z \geq 0$. But we also have
$b \gg a$ and hence $K(0) \approx 0$ to a good approximation.

An example of the fit for $N_T = 37$ and $M = 1.73\times10^6$ is graphed in
Figure~\ref{fig:rmse:a}.
The plot confirms visually, that the fit is good, with root-mean-square error
$\approx 1.46 \times 10^{-2}$.
In a Sco~X-1 search, these values of $N_T$ and $N_Q$ correspond to an
observation with
$\Tdrift = 10\,\mathrm{d}$ and $\Tobs = 370\,\mathrm{d}$, covering a bandwidth of
$N_Q \Delta
\fdrift = 1.0\,\mathrm{Hz}$.

Table~\ref{tab:c:rmse} presents $a$, $b$ and the root-mean-square error
of the fit for various practically motivated choices of $N_T$ and $N_Q$. The
error
is generally less than one per cent, giving confidence that~(\ref{eqn:kest})
is a good approximation.

When a signal is introduced (i.e. $\lambda > 0$), the situation is
complicated considerably, because the optimal path may travel through some
states containing the signal (drawn from a non-central chi-squared distribution)
and others containing noise only (drawn from a central chi-squared distribution).
We simplify things by considering the extreme case, where the
optimal path exactly matches the signal path. The simplification is conservative,
because in a real search it is possible
for the optimal path to include some noise-only bins yet still exceed the
threshold for
a detection.

In the extreme case, the cumulative distribution function for $\lambda > 0$ is
given by
\begin{align}
	K(z; \lambda) = 1 - Q_{k/2}(\sqrt{\lambda}, \sqrt{z})
\end{align}
after $k$ HMM steps,
where $Q_{k/2}$ is the Marcum-Q function,
\begin{align}
	Q_{k/2}(u, v) &= \frac{1}{u^{(k/2)-1}} \int_v^\infty
	\dee{x} x^{k/2} \\
	&\phantom{=} \exp[-(x^2 + u^2)/2] I_{(k/2)-1}(ux), \notag
\end{align}
and $I_{(k/2) - 1}$ is a modified Bessel function of order $(k/2) - 1$.

\subsection{Receiver operator characteristic (ROC) curve}
\label{ssec:c:farformula}
The HMM tracker determines the log likelihood $z$ that a given set of
parameters
corresponds to a signal. We choose a
threshold log likelihood $\zth$ and claim a detection
for $z > \zth$.
The false alarm probability, $\falsealarm$, quantifies how often pure noise
gives $z > \zth$,
causing a
spurious detection.
Given $\falsealarm$, we solve
\begin{align}
	K(\zth; \lambda = 0) = 1 - \falsealarm
\end{align}
for $\zth$.

Once $\falsealarm$ and hence $\zth$ are fixed,
some signals
by chance fail to be detected because they are too weak, relative
to the noise,
to produce $z > \zth$.
The
false dismissal probability, $\falsedismissal$, quantifies the probability of this
outcome.
Upon choosing $\falsedismissal$, we determine the weakest
signal that can be reliably detected by
solving
\begin{align}
	K(\zth; \lambda) = \falsedismissal
\end{align}
for $\lambda$ and hence $h_0$ via (\ref{eqn:lambdadef}).

\begin{figure*}
	\begin{subfigure}[b]{0.49\textwidth}
		\includegraphics[width=\textwidth]{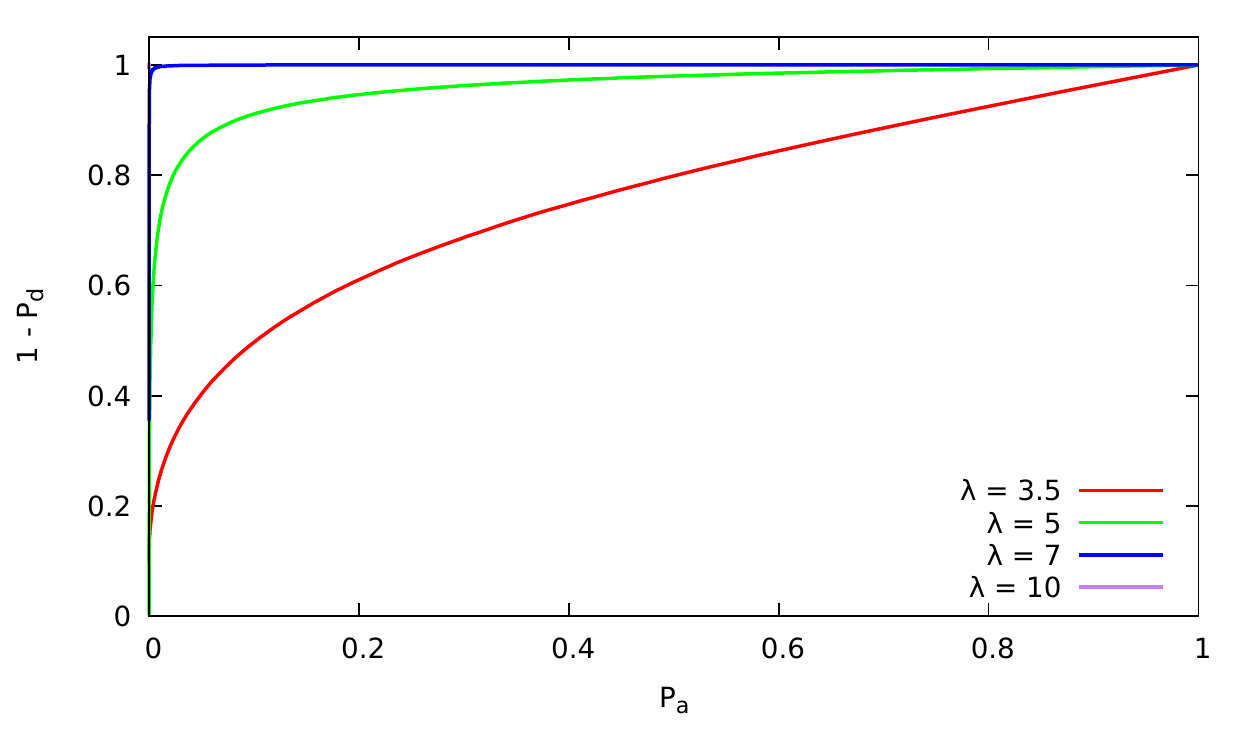}
		\caption{}
		\label{fig:roc}
	\end{subfigure}
	\hfill
	\begin{subfigure}[b]{0.49\textwidth}
		\includegraphics[width=\textwidth]{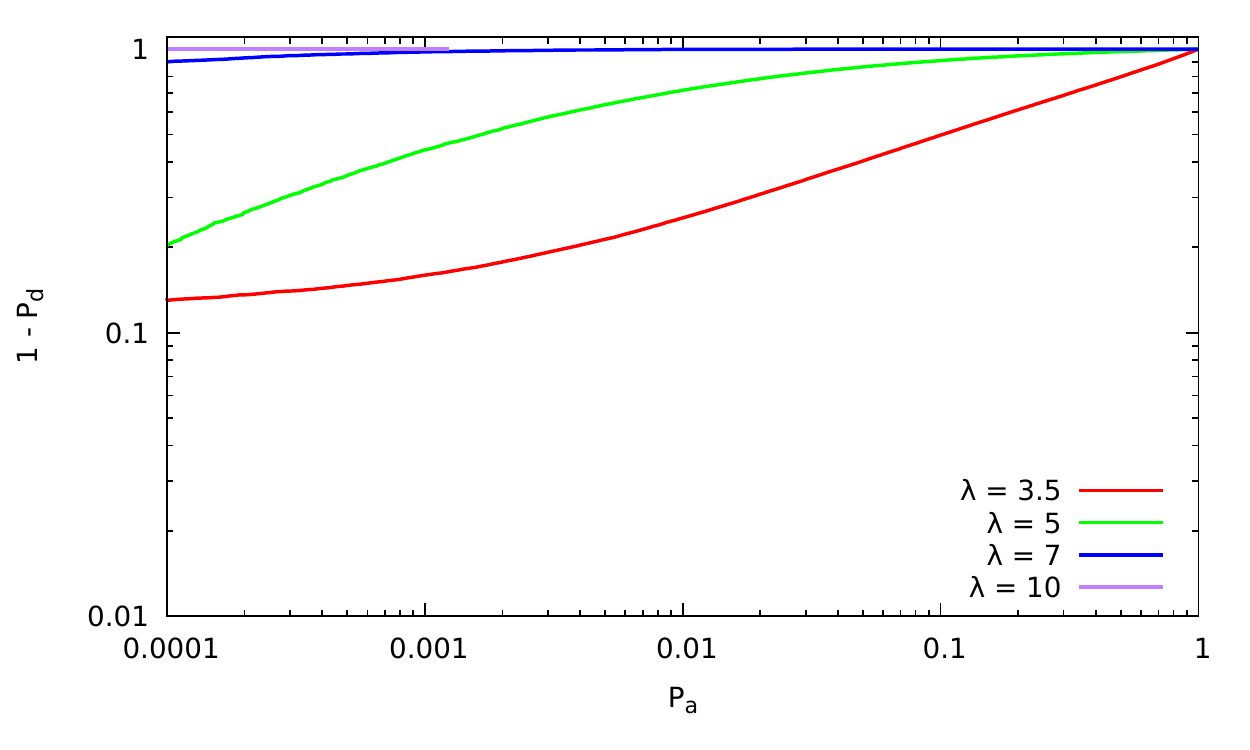}
		\caption{}
		\label{fig:roclog}
	\end{subfigure}
	\caption{Receiver operator characteristic curves
	for $N_T = 37$ blocks and
	four representative choices of
	$\lambda$, ranging from a strong signal ($\lambda = 10$,
	red curve) to a signal too weak to be reliably detected
	at the commonly used false alarm probability $\falsealarm = 0.1$
	($\lambda = 3.5$, purple curve).
	At each point along a curve, the
	vertical axis indicates the detection
	probability $1 - \falsedismissal$,
	and the horizontal axis indicates the false alarm probability $\falsealarm$.
	(a) Linear scale. (b) Log-log scale.
	Detection occurs
	when the optimal path exactly matches $f_\star(t)$, c.f. Viterbi score in
	Section~\ref{ssec:viterbiscore}.
	}
	\label{fig:rocall}
\end{figure*}

Figure~\ref{fig:roc}
displays ROC curves for four values of $\lambda$.
Each curve shows the tradeoff of $\falsealarm$ (on the horizontal axis) against
detection rate $1 - \falsedismissal$ (on the vertical axis).
The results are
replotted on logarithmic axes in
Figure~\ref{fig:roclog} to magnify the edges of the plot.
The
detection probability increases, as $\lambda$ increases.
It also rises superlinearly (linearly) with $\falsealarm$ for $\falsealarm \lesssim
0.1$ ($\falsealarm \gtrsim 0.1$).

\begin{figure}
	\includegraphics[width=0.49\textwidth]{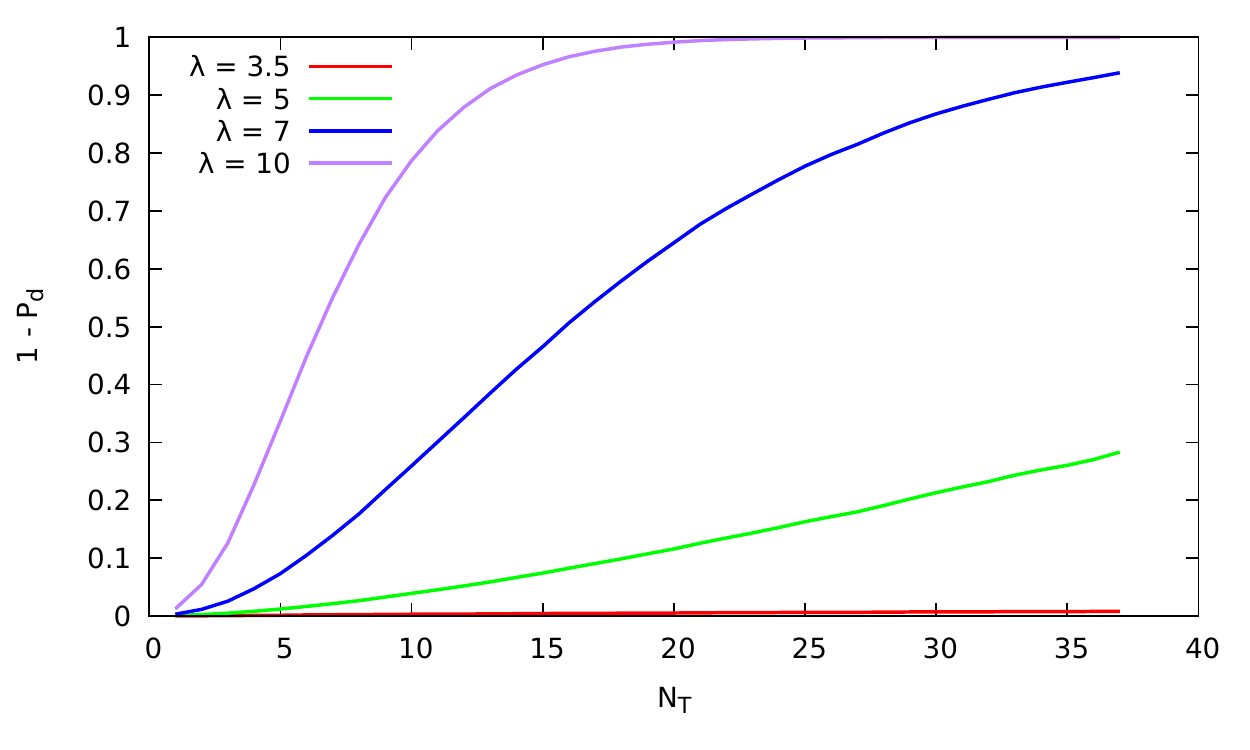}
	\caption{Detection probability versus number of HMM steps $N_T$ with
	$\Tdrift = 10\,\mathrm{d}$ for the same four values of $\lambda$ as in
	Figure~\ref{fig:rocall}. Detection occurs
	when the optimal path exactly matches $f_\star(t)$, c.f. Viterbi score in
	Section~\ref{ssec:viterbiscore}.}
	\label{fig:bstatkroc}
\end{figure}

Figure~\ref{fig:bstatkroc} shows
how the detection probability increases as more data
blocks are
processed, again for
the same four values of $\lambda$ as Figure~\ref{fig:rocall}.
As expected,
the detection probability rises, as $N_T$ and hence $\Tobs$ increase, keeping
$\Tdrift$ fixed.

\end{document}